\documentclass[useAMS,usenatbib]{mn2e}
\usepackage{txfonts}
\usepackage{graphicx} 
\usepackage{rotating}
\usepackage{caption}

\title[TempoNest: A Bayesian approach to pulsar timing analysis]{TempoNest: A Bayesian approach to pulsar timing analysis }
\author[L. Lentati et al.]{\parbox{\textwidth}{L. Lentati$^{1}$\thanks{E-mail:
ltl21@cam.ac.uk}, P. Alexander$^{1}$, M. P.  Hobson$^{1}$, F. Feroz$^{1}$, R. van Haasteren$^{2}$, K. J. Lee$^{3}$, R. M. Shannon$^{4}$ }\vspace{0.4cm}\\ %
$^{1}$Astrophysics Group, Cavendish Laboratory, JJ Thomson Avenue,  Cambridge, CB3 0HE, UK\\
$^{2}$Max-Planck-Institut fur Gravitationsphysik (Albert-Einstein-Institut), D-30167 Hannover, Germany\\
$^{3}$Max-Planck-Institut fur Radioastronomie, Auf dem Hugel 69, D-53121 Bonn, Germany\\
$^{4}$CSIRO Astronomy and Space Science, Australia Telescope National Facility, Box 76 Epping, NSW, 1710, Australia}

\begin{document}

\maketitle

\label{firstpage}

\begin{abstract}
A new Bayesian software package for the analysis of pulsar timing data is presented in the form of TempoNest which allows for the robust determination of the non-linear pulsar timing solution simultaneously with a range of additional stochastic parameters. This includes both red spin noise and dispersion measure variations using either power law descriptions of the noise, or through a model-independent method that parameterises the power at individual frequencies in the signal. 
We use TempoNest to show that at noise levels representative of current datasets in the European Pulsar Timing Array (EPTA) and International Pulsar Timing Array (IPTA) the linear timing model can underestimate the uncertainties of the timing solution by up to an order of magnitude. We also show how to perform Bayesian model selection between different sets of timing model and stochastic parameters, for example, by demonstrating that in the pulsar B1937+21 both the dispersion measure variations and spin noise in the data are optimally modelled by simple power laws.  Finally we show that not including the stochastic parameters simultaneously with the timing model can lead to unpredictable variation in the estimated uncertainties, compromising the robustness of the scientific results extracted from such analysis.

\end{abstract}


\section{Introduction}

The ever increasing precision of pulsar timing studies, combined, in particular, with the exceptional rotational stability of millisecond pulsars (MSPs) has resulted in a powerful tool for the pursuit of a wide range of scientific goals.  For example, in recent years pulsar timing has been used to find extrasolar planets \citep{2011Sci...333.1717B},  to study matter at super-nuclear densities in the interior of neutron stars \citep{2011MNRAS.414.1679E}, and the double pulsar system PSR J0737-3039A/B, provides precise measurements of several `post Keplerian' parameters allowing for stringent tests of general relativity \citep{2006Sci...314...97K}.

For a detailed review of pulsar timing refer to e.g. \cite{2004hpa..book.....L}.  In brief, the arrival times of pulses (TOAs) for a particular pulsar will be recorded by an observatory in a series of discrete observations over a period of time.  These arrival times must all be transformed into a common frame of reference, the solar system barycenter, in order to correct for the motion of the Earth.

A model for the pulsar can then be fit to the TOAs that characterises the properties of the pulsar's orbital motion, as well as its timing properties such as its orbital frequency and spin down.  This is most commonly carried out using the TEMPO\footnote{http://www.atnf.csiro.au/research/pulsar/tempo}, and more recently, TEMPO2 pulsar-timing packages  \citep{2006MNRAS.369..655H, 2006MNRAS.372.1549E, 2009MNRAS.394.1945H}. TEMPO2 uses an initial guess to the timing model to generate a set of pre-fit residuals.  A Fisher-matrix approximation to the timing model parameters is then calculated and a linear least-squares method is utilised to improve the fit.  If desired, multiple iterations can be performed such that the best-fit values for the timing model from the previous iteration are used as the starting guess for the next, until convergence is reached.

When performing this fitting process, TEMPO2 considers the TOAs to be solely the sum of a deterministic signal due to the timing model, and a white noise component described completely by the TOA uncertainties.  In realistic datasets however this assumption is rarely true.  If additional stochastic processes such as intrinsic red spin noise due to rotational irregularities in the neutron star \citep{2010ApJ...725.1607S} or correlated noise due to a stochastic gravitational wave background (GWB) generated by, for example, coalescing black holes (e.g. \citealt{2003ApJ...583..616J,2001astro.ph..8028P}) are present in the data then power from these contributions will be absorbed by the timing model, affecting the accuracy of the parameter estimation.  

Recently, \cite{2011MNRAS.418..561C} (henceforth C11) proposed a method of improving the timing model fit by using the Cholesky-decomposition of the covariance matrix describing the properties of these additional stochastic processes in the TOAs, calculated from the power spectral density of the timing residuals.  This can be used to whiten the residuals, after which parameter estimation is performed in these transformed observations using ordinary least squares.

In \cite{2013MNRAS.428.1147V} (henceforth vHL2013), however, it is shown that after fitting for the timing model the resulting residuals are not time stationary, and as such the power spectral density of those residuals is not a well defined mathematical quantity.  In addition because the method in C11 does not account for the covariance between the timing model and the stochastic processes, the uncertainties associated with the parameter estimates, in particular those associated with the quadratic spin down, are not optimal.  The preferred approach is therefore to perform a joint analysis of the deterministic timing model and any additional stochastic components present as in vHL2013.

However, when performing a Bayesian analysis using the linearised timing model as presented in vHL2013 it is not possible to perform model selection between different sets of timing model parameters using the evidence.  This is because the maximum likelihood value at which the linearisation is performed will depend upon the exact set of model parameters included, and as such, both the data and the model will vary as the parameter space changes.  It is also not clear how the estimation of the uncertainties of the timing model parameters depends on the linearisation process, especially in the regime where the signal to noise ratio might be low, and the Fisher-matrix approximation will be poor.

In this paper we present a solution to these problems in the form of TempoNest.  TempoNest provides a means of performing a simultaneous analysis of either the linear or non-linear timing model and additional stochastic parameters using the Bayesian inference tool MultiNest (\citealt{2008MNRAS.384..449F, 2009MNRAS.398.1601F}) to efficiently explore this joint parameter space, whilst using TEMPO2 as an established means of evaluating the timing model at each point in that space.  TempoNest allows for robust model selection between different sets of timing model or noise parameters, and requires only basic prior knowledge of the timing model.

In Section \ref{Section:Bayes} we will describe the basic principles of our Bayesian approach to data analysis, giving a brief overview of how it may be used to perform model selection, and introduce MultiNest.  In Section \ref{Section:Models} we will describe the stochastic models currently available for use in TempoNest to include with the timing model, including the white noise modifiers EFAC and EQUAD, along with descriptions of both red spin noise and dispersion measure variations.

We will then perform a series of tests using TempoNest designed to show some of the included functionality.  In Section \ref{Section:Sims} we use simulated data to compare the non-linear and linear approximation to the timing model across different noise regimes designed to represent both future and current datasets, whilst in Section \ref{Section:RealData} we apply TempoNest to two sets of publicly available data, firstly of the binary pulsar B1855+09 and then the isolated pulsar B1937+21.  We show how TempoNest can be used to perform Bayesian model selection between different sets of timing model and stochastic parameters, and for the latter case, also compare the parameters estimates and uncertainties for the timing solutions produced by TempoNest, Tempo2, and the SpectralModel plug-in for Tempo2 that uses the principles described in C11.

We note that the aim of this paper is not to provide a user manual for TempoNest, but rather give an overview of its functionality.  A development build of TempoNest is currently available online  \footnote{https://github.com/LindleyLentati/TempoNest}, with a full public release planned in the near future.

This research is the result of the common effort to directly detect gravitational waves using pulsar timing, known as the European Pulsar Timing Array (EPTA) \citep{2008AIPC..983..633J} \footnote{www.epta.eu.org/}.

\section[]{Bayesian Inference}
\label{Section:Bayes}

Our method for performing pulsar timing analysis is built upon the principles of Bayesian inference, which provides a consistent approach to the estimation of a set of parameters $\Theta$ in a model or hypothesis $H$ given the data, $D$.  Bayes' theorem states that:

\begin{equation}
\mathrm{Pr}(\Theta \mid D, H) = \frac{\mathrm{Pr}(D\mid \Theta, H)\mathrm{Pr}(\Theta \mid H)}{\mathrm{Pr}(D \mid H)},
\end{equation}
where $\mathrm{Pr}(\Theta \mid D, H) \equiv \mathrm{Pr}(\Theta)$ is the posterior probability distribution of the parameters,  $\mathrm{Pr}(D\mid \Theta, H) \equiv L(\Theta)$ is the likelihood, $\mathrm{Pr}(\Theta \mid H) \equiv \pi(\Theta)$ is the prior probability distribution, and $\mathrm{Pr}(D \mid H) \equiv Z$ is the Bayesian Evidence.

In parameter estimation, the normalizing evidence factor is usually ignored, since it is independent of the parameters $\Theta$.   Inferences are therefore obtained by taking samples from the (unnormalised) posterior using, for example, standard Markov chain Monte Carlo (MCMC) sampling methods.  

In contrast to parameter estimation, for model selection the evidence takes the central role and is simply the factor required to normalize the posterior over $\Theta$:

\begin{equation}
Z = \int L(\Theta)\pi(\Theta) \mathrm{d}^n\Theta,
\label{eq:Evidence}
\end{equation}
where $n$ is the dimensionality of the parameter space.  

As the average of the likelihood over the prior, the evidence is larger for a model if more of its parameter space is likely and smaller for a model where large areas of its parameter space have low likelihood values, even if the likelihood function is very highly peaked.  Thus, the evidence automatically implements Occam's razor: a simpler theory with a compact parameter space will have a larger evidence than a more complicated one, unless the latter is significantly better at explaining the data.  

The question of model selection between two models $H_0$ and $H_1$ can then be decided by comparing their respective posterior probabilities, given the observed data set $D$, via the model selection ratio $R$:

\begin{equation}
R= \frac{P(H_1\mid D)}{P(H_0\mid D)} = \frac{P(D \mid H_1)P(H_1)}{P(D\mid H_0)P(H_0)} = \frac{Z_1}{Z_0}\frac{P(H_1)}{P(H_0)},
\label{Eq:Rval}
\end{equation}
where $P(H_1)/P(H_0)$ is the a priori probability ratio for the two models, which can often be set to unity but occasionally requires further consideration.

\subsection{Nested Sampling and evaluating the evidence}

Evaluation of the multidimensional integral in Eq. \ref{eq:Evidence} is a challenging numerical task.  Standard techniques like thermodynamic integration \citep{Thermo} are extremely computationally expensive, which makes evidence evaluation  at least an order-of-magnitude more costly than parameter estimation.  Some fast approximate methods have been used for evidence evaluation, such as treating the posterior as a multivariate Gaussian centered at its peak (see e.g. \citealt{2002MNRAS.335..377H} ), but this approximation is clearly a poor one for multimodal posteriors (except perhaps if one performs a separate Gaussian approximation at each mode).  The Savage-Dickey density ratio has also been proposed (see e.g. \citealt{2007MNRAS.378...72T}) as an exact, and potentially faster, means of evaluating evidences, but is restricted to the special case of nested hypotheses and a separable prior on the model parameters.

The nested sampling approach \citep{2004AIPC..735..395S}  is a Monte-Carlo method targeted at the efficient calculation of the evidence, but also produces posterior inferences as a by-product.  

Nested sampling considers the prior volume $X$ where the likelihood is greater than some value $\lambda$, which can be written as:

\begin{equation}
X(\lambda) = \int_{L(\Theta)>\lambda}  \pi(\Theta) \mathrm{d}^n\Theta.
\end{equation}
This allows us to rewrite Eq. \ref{eq:Evidence} as a one-dimensional integral over $\lambda$:
\begin{equation}
Z = \int_0^{\infty}  X(\lambda) \mathrm{d}\lambda.
\end{equation}
When the inverse of $X(\lambda)$, the likelihood value that corresponds to a given prior volume, $L(X)$, exists this integral can then be written:
\begin{equation}
Z = \int_0^1  L(X) \mathrm{d}X,
\end{equation}
and so the evidence can be calculated as the weighted sum of a set of $M$  values of $X$:
\begin{equation}
Z=\sum_{i=1}^M L_iw_i.
\end{equation}
where the weights $w_i$ are simply given by the trapezium rule $w_i = \frac{1}{2}\left(X_{i-1} - X_{i+1} \right)$.

\subsection{MultiNest}

In \cite{2009MNRAS.398.1601F} and \cite{2008MNRAS.384..449F} this nested sampling framework was built upon with the introduction of the MULTINEST algorithm, which provides an efficient means of sampling from posteriors that may contain multiple modes and/or large (curving) degeneracies, and also calculates the evidence.  Since its release MULTINEST has been used successfully in a wide range of astrophysical problems, from detecting the Sunyaev-Zel'dovich effect in galaxy clusters \citep{2012arXiv1210.7771C}, to inferring the properties of a potential stochastic gravitational wave background in pulsar timing array data \citep{2013PhRvD..87j4021L} (henceforth L13).

In brief, the MultiNest algorithm operates by first drawing a set of $N_{\mathrm{live}}$ points from the prior $\pi(\Theta)$.  An ellipsoidal decomposition is then performed such that the full set of live points is contained within a set of ellipsoids.  At each subsequent iteration $i$ a point is drawn with likelihood $L$ from the union of these ellipsoids and is checked to see if it satisfies the constraint $L > L_i$ where $L_i$ is the lowest likelihood value present in the set of live points at that iteration.  If this constraint is satisfied the point replaces the lowest likelihood point in the live set with a probability $1/n_e$ where $n_e$ is the number of ellipsoids in which the new point lies.

In high dimensions most of the volume in the ellipsoids lies in their outer shells, thus when the decomposition extends beyond the true iso-likelihood surface, the acceptance rate of new points can decrease significantly.  In order to maintain high sampling efficiency in high dimensions MultiNest therefore contains a `constant efficiency' mode.  Here the total volume enclosed by the ellipsoids is adjusted such that the sampling efficiency meets some user set target.  However whilst this mode is adequate for parameter estimation, the evidence values are not reliable. 

Recently, however, the MULTINEST algorithm has been updated to include the concept of importance nested sampling \citep{2013arXiv1301.6450C} (INS) which provides a solution to this problem.  Full details can be found in \cite{2013arXiv1306.2144F}, but the key difference is that, where with normal nested sampling the rejected points play no further role in the sampling process, INS uses every point sampled to contribute towards the evidence calculation.  One outcome of this approach is that even when running in constant efficiency mode the evidence calculated is reliable even in higher ($\sim$ 50) dimensional problems.

In pulsar timing analysis we will often have to deal with timing models that can contain $> 20$ parameters, which, when combined with the properties of the stochastic component of the signal can result in a total dimensionality of 50-60.  As such, the ability to run in constant efficiency mode whilst still obtaining accurate values for the evidence when these higher dimensional problems arise is crucial in order to perform reliable model selection.   

\section{Pulsar timing likelihood}
\label{Section:Models}

For any pulsar we can write the TOAs for the pulses as a sum of both a deterministic and a stochastic component:

\begin{equation}
\mathbf{t}_{\mathrm{tot}} = \mathbf{t}_{\mathrm{det}} + \mathbf{t}_{\mathrm{sto}},
\end{equation}
where $\mathbf{t}_{\mathrm{tot}}$ represents the $n$ TOAs for a single pulsar, with $\mathbf{t}_{\mathrm{det}}$ and $\mathbf{t}_{\mathrm{sto}}$ the deterministic and stochastic contributions to the total respectively, where any contributions to the latter will be modelled as random Gaussian processes.  
Writing the deterministic signal due to the timing model as $\bmath{\tau}(\bmath{\epsilon})$, and the uncertainty associated with a particular TOA $i$ as $\sigma_i$ we can write the likelihood that the data is described solely by the timing model as:

\begin{equation}
\label{Eq:TempoLike}
\mathrm{Pr}(\mathbf{t} | \bmath{\epsilon}) \propto \left(\prod_{i=1}^n\sigma_i^2\right)^{-\frac{1}{2}}\exp{\left(-\frac{1}{2}\sum_{i=1}^n\frac{(t_i - \tau(\bmath{\epsilon})_i)^2}{\sigma_i^2}\right)}.
\end{equation}
This represents the simplest model choice possible in TempoNest, including only those free parameters present in the TEMPO2 fit.  From here we can now begin to make our model for the stochastic contribution to the signal more realistic by introducing additional parameters to describe the white and red noise components, in order to compare the evidence with this simpler model and determine the optimal set of parameters supported by the data.

\subsection{Additional white noise}
\label{Section:White}

When dealing with pulsar timing data, the properties of the white noise can be separated into two components:

\begin{description}
  \item[1:] For a given pulsar, each TOA has an associated error bar, the size of which will vary across a set of observations.  We can introduce an extra free parameter, denoted EFAC, to account for possible mis-calibration of this radiometer noise \citep{2006MNRAS.369..655H}.  The EFAC parameter therefore acts as a multiplier for all the TOA error bars for a given pulsar, observed with a particular system. TempoNest allows for either a single EFAC parameter to be estimated for all TOAs for a given pulsar, or, where the observing system has been flagged for each TOA, a separate EFAC can be included for each system.\\
 \item[2:] A second white noise component is also used to represent some additional source of time independent noise, which we call EQUAD.  In principal this parameter represents something physical about the pulsar, the high frequency tail of the pulsar's red spin noise power spectrum, or, `jitter' \citep{2012MNRAS.420..361L}, and so should be independent of the observing system.  Differences in the integration times between TOAs for different observing epochs can muddy this physical interpretation however, and as such as with EFAC, either a single EQUAD parameter can be estimated for all TOAs for a given pulsar, or for each flagged system separately.
\end{description} 
We can therefore rewrite the error $\sigma_i$ associated with each TOA $i$ as $\hat{\sigma}_i$ so that:
\begin{equation}
\hat{\sigma}_i^2 = (\alpha_i\sigma_i)^2 + \beta_i^2
\end{equation}
where $\alpha$ and $\beta$ represent the EFAC and EQUAD parameters applied to TOA $i$ respectively.   Note this is not how Tempo2 defines the relationship between the EQUAD and EFAC parameters.  Thus Eq. \ref{Eq:TempoLike} can be trivially rewritten to include the new white noise parameters as:

\begin{equation}
\label{Eq:WhiteLike}
\mathrm{Pr}(\mathbf{t} | \bmath{\epsilon}, \bmath{\alpha}, \beta) \propto \left(\prod_{i=1}^n\hat{\sigma}_i^2\right)^{-\frac{1}{2}}\exp{\left(-\frac{1}{2}\sum_{i=1}^n\frac{(t_i - \tau(\bmath{\epsilon})_i)^2}{\hat{\sigma}_i^2}\right)}.
\end{equation}

\subsection{Additional red noise}

TempoNest currently supports two methods for describing the intrinsic red noise properties of the pulsar, the recently introduced model independent frequency domain method described in L13 and the power law model, time domain method described in \citep{2009MNRAS.395.1005V} (henceforth vH2009).

\subsubsection{L13 method}
\label{Section:RedL}

We begin by writing the red noise component of the stochastic signal, which we will denote $\mathbf{t}_{\mathrm{red}}$, in terms of its Fourier coefficients $\mathbf{a}$ so that $\mathbf{t}_{\mathrm{red}} = \mathbf{F}\mathbf{a}$ where $\mathbf{F}$ denotes the Fourier transform such that for frequency $\nu$ and time $t$ we will have both:

\begin{equation}
F_{\nu,t} = \frac{1}{T}\sin\left(2\pi\nu t\right),
\end{equation}
and an equivalent cosine term.  Here $T$ represents the total observing span for the pulsar, and $\nu$ the frequency of the signal to be sampled.  Defining the number of coefficients to be sampled by $n_{\mathrm{max}}$, TempoNest will then include the set of frequencies with values $n/T$, where $n$ extends from 1 to $n_{\mathrm{max}}$.
For typical PTA data \cite{2012MNRAS.423.2642L} show that a low frequency cut off of 1/T is sufficient to accurately describe the expected long term variations present in the data. If necessary though it is also possible to specify arbitrary sets of frequencies such that terms with ν << 1/T can be included in the model, or to allow noise terms where the frequency itself is a free parameter. 

For a single pulsar the covariance matrix $\bmath{\varphi}$ of the Fourier coefficients $\mathbf{a}$ will be diagonal, with components

\begin{equation}
\label{Eq:BPrior}
\varphi_{ij} = \left< a_ia_j^*\right> = \varphi_{i}\delta_{ij},
\end{equation}
where there is no sum over $i$, and the set of coefficients $\{\varphi_{i}\}$ represent the theoretical power spectrum for the residuals.  

As discussed in L13, whilst Eq \ref{Eq:BPrior} states that the Fourier modes are orthogonal to one another, this does not mean that we assume they are orthogonal in the time domain where they are sampled, and it can be shown that this non-orthogonality is accounted for within the likelihood.  Instead, in Bayesian terms, Eq. \ref{Eq:BPrior} represents our prior knowledge of the power spectrum coefficients within the data.  We are therefore stating that, whilst we do not know the form the power spectrum will take, we know that the underlying Fourier modes are still orthogonal by definition, regardless of how they are sampled in the time domain.  It is here then that, should one wish to fit a specific model to the power spectrum coefficients at the point of sampling, such as a broken, or single power law, the set of coefficients $\{\varphi_{i}\}$ should be given by some function $f(\Theta)$, where we sample from the parameters $\Theta$ from which the power spectrum coefficients $\{\varphi_{i}\}$ can then be derived.

We can then write the joint probability density of the timing model, white noise parameters, power spectrum coefficients and the signal realisation, Pr$(\bmath{\epsilon}, \bmath{\alpha}, \beta, \{\varphi_i\}, \mathbf{a} \;|\; \mathbf{t})$, as:

\begin{eqnarray}
\label{Eq:Prob}
\mathrm{Pr}(\bmath{\epsilon}, \bmath{\alpha}, \beta, \{\varphi_i\}, \mathbf{a} \;|\; \mathbf{t}) \; &\propto& \; \mathrm{Pr}(\mathbf{t} |  \bmath{\epsilon}, \bmath{\alpha}, \beta, \mathbf{a}) \; \\\ \nonumber
&\times & \mathrm{Pr}(\mathbf{a} | \{\varphi_i\}) \; \mathrm{Pr}(\{\varphi_i\}). \nonumber
\end{eqnarray}
For our choice of $\mathrm{Pr}(\{\varphi_i\})$ we use an uninformative prior that is uniform in $\log_{10}$ space, and draw our samples from the parameter $\rho_i = \log_{10}(\varphi_i)$ instead of $\varphi_i$ which has the added advantage that we avoid unnecessary rejections due to samples that have negative coefficients in the sampling process.  Given this choice of prior the conditional distributions that make up Eq. \ref{Eq:Prob} can be written:

\begin{eqnarray}
\label{Eq:ProbTime}
& &\mathrm{Pr}(\mathbf{t} |\bmath{\epsilon}, \bmath{\alpha}, \beta, \mathbf{a}) \; \propto \; \frac{1}{\sqrt{\mathrm{det}(\mathbf{N})}}  \\
& \times & \exp\left[-\frac{1}{2}(\mathbf{t} -  \bmath{\tau}(\bmath{\epsilon}) - \mathbf{F}\mathbf{a})^T\mathbf{N}^{-1}(\mathbf{t} -  \bmath{\tau}(\bmath{\epsilon}) - \mathbf{F}\mathbf{a})\right] \nonumber
\end{eqnarray}
where $\mathbf{N}_{ij} = \hat{\sigma}^2_{i}\delta_{ij}$ and represents the white noise errors in the residuals and:

\begin{equation}
\label{Eq:ProbFreq}
\mathrm{Pr}(\mathbf{a} | \{\rho_i\}) \; \propto \; \frac{1}{\sqrt{\mathrm{det}\bmath{\varphi}}} \exp\left[-\frac{1}{2}\mathbf{a}^{*T}\bmath{\varphi}^{-1}\mathbf{a}\right].
\end{equation}
In TempoNest we then marginalise over all Fourier coefficients $\mathbf{a}$ analytically in order to find the posterior for the remaining parameters alone.

When performing this marginalisation we use a uniform prior for the Fourier coefficients, so that, denoting $\mathbf{t}  -  \bmath{\tau}(\bmath{\epsilon})$ as $\bmath{\delta t}$,  $(\mathbf{F}^T\mathbf{N}^{-1}\mathbf{F} + \bmath{\varphi}^{-1})$ as $\bmath{\Sigma}$ and $\mathbf{F}^T\mathbf{N}^{-1}\bmath{\delta t}$ as $\mathbf{d}$ our marginalised posterior is given by:

\begin{eqnarray}
\label{Eq:Margin}
\mathrm{Pr}(\bmath{\epsilon}, \bmath{\alpha}, \beta, \{\varphi_i\} | \mathbf{t}) &\propto& \frac{\mathrm{det} \left(\mathbf{\Sigma}\right)^{-\frac{1}{2}}}{\sqrt{\mathrm{det} \left(\bmath{\varphi}\right)~\mathrm{det}\left(\mathbf{N}\right)}} \\
&\times&\exp\left[-\frac{1}{2}\left(\bmath{\delta t}^T\mathbf{N}^{-1} \bmath{\delta t} - \mathbf{d}^T\mathbf{\Sigma}^{-1}\mathbf{d}\right)\right]. \nonumber
\end{eqnarray}

\subsubsection{vH2013 method}
\label{Section:vHRed}
Here we begin by parameterising the red noise process using a power law spectral density of the form:

\begin{equation}
S(f) = A^2\left(\frac{f}{1\mathrm{yr}^{-1}}\right)^{\gamma},
\end{equation}
where $S(f)$ is the spectral density at frequency $f$, $A$ is the amplitude of the red noise process and $\gamma$ is the spectral index.
We write the time domain covariance matrix $\mathbf{C}^{Red}_{ij}$ between observations $i$ and $j$ as given in vH2009:

\begin{eqnarray}
\label{Eq:Red}
\mathbf{C}^{Red}_{ij}  &=& \frac{A^2}{f_L^{\gamma-1}}\left\{\Gamma(1-\gamma)\sin\left(\frac{\pi\gamma}{2}\right)\right.\\
&\times&\left.(f_L\tau)^{\gamma-1} - \sum_{n=0}^{\infty}(-1)^n\frac{(f_l\tau)^{2n}}{(2n)!(2n+1-\gamma)}\right\}. \nonumber
\end{eqnarray}
where $f_L$ is a low frequency cut off and $\tau = 2\pi(t_{i}-t_{j})$ with $t_{i}$ the $i$th TOA.  In \citep{2009MNRAS.395.1005V} it is shown that the quadratic spin down acts to absorb any contribution to the signal that arises from the choice of this low frequency cut off, and as such it is only necessary to choose $f_L$ so that $1/f_L$ is much greater than the observing time span in order to obtain rapid convergence of the expression.   The result of this however, is that the level of uncertainty in the spin down parameters will be affected to a much greater extent than any others by the presence of low frequency stochastic processes in the data, a fact that we will return to in Section \ref{Section:1937}.
Finally denoting the white noise covariance matrix  $\mathbf{N}$ as before we can write the total covariance matrix describing our simulated residuals $\mathbf{C}_{ij}$ as:

\begin{equation}
 \mathbf{C} = \mathbf{C}^{Red} + \mathbf{N}.
 \end{equation}
We can then write our likelihood as:

\begin{eqnarray}
\label{Eq:Rlike}
\mathrm{Pr}(\mathbf{t} | \bmath{\epsilon},\bmath{\alpha}, \beta, A, \gamma)& = & \frac{1}{\sqrt{(2\pi)^n\mathrm{det}\mathbf{C}}} \\
&\times& \exp{\left(-\frac{1}{2}(\mathbf{t} -  \bmath{\tau}(\bmath{\epsilon}))^T\mathbf{C}^{-1}(\mathbf{t} -  \bmath{\tau}(\bmath{\epsilon}))\right)}. \nonumber
\end{eqnarray}

\subsection{Including dispersion measure variations}

The plasma located in the interstellar medium (ISM), as well as in solar winds and the ionosphere can result in delays in the propagation of the pulse signal between the pulsar and the observatory, an effect that appears as a red noise signal in the timing residuals.  

Unlike other red noise signals however, the severity of the observed dispersion measure variations is dependant upon the observing frequency, and as such we can use this additional information to isolate the component of the red noise that results from this effect.

In particular, the group delay $t_g(\nu)$ for a frequency $\nu$ is given by the relation:

\begin{equation}
t_g(\nu) = DM/(K\nu^2)
\end{equation}
where the dispersion constant $K$ is given by:

\begin{equation}
K \equiv 2.41 \times 10^{-16}~\mathrm{Hz^{-2}~cm^{-3}~pc~s^{-1}}
\end{equation}
and the dispersion measure is defined as the integral of the electron density $n_e$ from the Earth to the pulsar:

\begin{equation}
DM = \int_0^L n_e \mathrm{d}l.
\end{equation}

Dispersion measure corrections can be included in the analysis as an additional set of stochastic parameters with only minor modifications to the equations \ref{Eq:Margin} and \ref{Eq:Rlike} allowing as before, using either a power law model or the model independent description.  In both cases we begin by first defining a vector $\bmath{D}$ of length equal to the number of observations for a given pulsar as:

\begin{equation}
D_i = 1/(K\nu^2_i)
\end{equation}
for observation $i$ with observing frequency $\nu_i$.

\subsubsection{Model independent method}

For the model independent approach we then need to make a change to our basis vectors such that our dispersion measure Fourier modes are described by:

\begin{equation}
F^{DM}_{\nu,t_i} = \frac{1}{T}\sin\left(2\pi\nu_s t_i\right)D_i
\end{equation}
and an equivalent cosine term, where $T$ is the length of the observing timespan, and $\nu_s$ now explicitly denotes the frequency of the signal to be parameterised as before, where the set of frequencies to be included is defined in the same way as for the red spin noise.  Unlike when modelling the red spin noise, we no longer have the quadratic in the timing model to act as a proxy to the low frequency ($\nu_s < 1/T$) DM variations in our data.  As such these terms must be accounted for either by explicitly including these low frequencies in the model, or by including a quadratic in DM to act as a proxy, as with the red noise, defined as: 

\begin{equation}
 Q_{\mathrm{DM}}(t_i)= \alpha_0 t_iD_i + \alpha_1 t_i^2D_i
\end{equation}
with $\alpha_{0,1}$ free parameters to be fit for, and $t_i$ the barycentric arrival time for TOA $i$.  This can be achieved most simply be adding the timing model parameters $DM1$ and $DM2$ into the pulsar parameter file, and allowing TempoNest to include them in the fit.

\subsubsection{Time domain power law model}

For a detailed discussion of this approach, and comparisons to existing methods see Lee et al. (submitted).  In brief, we transform our red noise covariance matrix $C^{Red}_{ij}$ to:

\begin{equation}
C^{DM}_{ij} = C^{Red}_{ij}D_iD_j.
\end{equation}
The total noise covariance matrix can therefore be rewritten as:
\begin{equation}
\mathbf{C}^{tot} = \mathbf{C}^{Red} + \mathbf{N} + \mathbf{C}^{DM}.
\end{equation}

\section{Analytical marginalisation over the timing model}
\label{Section:Margin}

Despite having the ability to fit simultaneously for all the timing model parameters and the stochastic properties of the noise present in the signal, there may be times where it is preferable to marginalise over some of the timing model parameters analytically in order to decrease the dimensionality of the problem.  For example, a set of TOAs for a single pulsar might be the combination of many different sets of observations taken by different observatories, with phase jumps fitted between each set.  If the specific values of these jumps are not of interest then the analysis might be performed faster if the decrease in the number of calculations required to explore the smaller dimensional  space outweighs the increase in the calculation time that results from the matrix operations required by the marginalisation process.

If we separate the timing model into a contribution from the set of parameters that we wish to parameterise $\bmath{\tau}(\bmath{\epsilon})$ and a contribution from the set of $m$ parameters that we plan to marginalise over analytically $\bmath{\tau}(\bmath{\epsilon'})$  then we can write the probability that the data $\mathbf{t}$ is described by the remaining parameters $\bmath{\epsilon}$ and any additional parameters $\bmath{\theta}$ we wish to include as:

\begin{eqnarray}
\mathrm{Pr}(\mathbf{t} | \bmath{\epsilon}, \bmath{\theta}) &=& \int \; \mathrm{d}^m\bmath{\epsilon'} \; \mathrm{Pr}(\bmath{\epsilon'}) \;\mathrm{Pr}(\mathbf{t} | \bmath{\epsilon'},\bmath{\epsilon},\bmath{\theta})
\end{eqnarray}
Using a uniform prior on the $m$ $\bmath{\epsilon'}$  parameters, we use the same approach as described in  \citep{2013MNRAS.428.1147V} to perform this marginalisation process analytically.  This results in a  set of equations \ref{Eq:WhiteLikeMargin},  \ref{Eq:LLikeMargin} that exist in parallel to Eqns \ref{Eq:Margin} and \ref{Eq:Rlike} :

\begin{eqnarray}
\label{Eq:WhiteLikeMargin}
\mathrm{Pr}(\mathbf{t} | \bmath{\epsilon}, \bmath{\alpha}, \beta) &=& \frac{1}{\sqrt{(2\pi)^{n-m}\mathrm{det}(\mathbf{G}^T\mathbf{CG})}}  \\ 
& \times& \exp{\left(-\frac{1}{2}(\bmath{\delta t})^T\mathbf{G}(\mathbf{G}^T\mathbf{CG})^{-1}\mathbf{G}^T(\bmath{\delta t})\right)} \nonumber.
\end{eqnarray}
where $\bmath{\delta t} = \mathbf{t} - \bmath{\tau}(\bmath{\epsilon})$,  $\mathbf{C}$ is the $n\times n$ noise covariance matrix as before and $\mathbf{G}$ is the $n \times (n-m)$ matrix that performs the marginalisation whose derivation will not be given here but is described in \citep{2013MNRAS.428.1147V},

\begin{eqnarray}
\label{Eq:LLikeMargin}
\mathrm{Pr}(\bmath{\epsilon}, \bmath{\alpha}, \beta, \{\varphi_i\} | \mathbf{t}) &\propto& \frac{\mathrm{det} \left(\mathbf{\hat{\Sigma}}\right)^{-\frac{1}{2}}}{\sqrt{\mathrm{det} \left(\bmath{\varphi}\right)~\mathrm{det}\left(\mathbf{\hat{N}}\right)}} \\
&\times&\exp\left[-\frac{1}{2}\left(\bmath{\delta t}^T\mathbf{\mathbf{\hat{N}}}^{-1} \bmath{\delta t} - \mathbf{\hat{d}}^T\mathbf{\hat{\Sigma}}^{-1}\mathbf{\hat{d}}\right)\right], \nonumber
\end{eqnarray}
where $\mathbf{\hat{N}} = \mathbf{G}(\mathbf{G}^T\mathbf{NG})^{-1}\mathbf{G}^T$, $\mathbf{\hat{\Sigma}} = (\mathbf{F}^T{\mathbf{\hat{N}}}^{-1}\mathbf{F} + \bmath{\varphi}^{-1})$ and $\mathbf{\hat{d}} = \mathbf{F}^T\mathbf{\hat{N}}^{-1}\bmath{\delta t}$.

\section{Linear approximation to the timing model}

We would like to compare the results of the non-linear analysis of the timing model afforded by TempoNest, with those that can be obtained from the linear approximation, and so we provide a brief description of the linear model below.  

Given an initial estimate of the $m$ timing model parameters $\epsilon_{\mathrm{0}i}$ a linear approximation can be performed such that any deviations from that estimate are encapsulated using the $m$ parameters $\delta\epsilon_i$ such that:

\begin{equation}
\delta\epsilon_i = \epsilon_i- \epsilon_{\mathrm{0}i}.
\end{equation}
Therefore, writing the set of post--fit residuals that results from the subtraction of the initial estimate of the timing model from our TOAs as  $\mathbf{\delta t}_{\mathrm{post}}$ we can express the change in these residuals that results from the deviation in the timing model parameters $\bmath{\delta\epsilon}$ as: 
\begin{equation}
\mathbf{\delta t} = \mathbf{\delta t}_{\mathrm{post}} -  \mathrm{\bf{M}}\bmath{\delta\epsilon},
\end{equation}
where $\mathbf{M}$ is the $n\times m$ `design matrix' which describes the dependence of the timing residuals on the model parameters.  

Therefore in all previous equations, we can simply substitute $ \mathbf{t} - \bmath{\tau}(\bmath{\epsilon})$ for $ \mathbf{\delta t}_{\mathrm{post}} -  \mathrm{\bf{M}}\bmath{\delta\epsilon}$ in order to evaluate the linear approximation to the timing model.

\section{Application to simulated data}
\label{Section:Sims}

In order to compare the parameter estimates obtained through both the non-linear and the linearised timing models we use a series of three simulations, details of which are given below.  The simulations are designed to make it progressively more difficult to extract the correct timing model parameters, due both to increasing the amplitude of the white noise in the data, and increasing the complexity of the noise by including additional red noise signals.

\begin{description}
  \item[Simulation 1:] The TOAs consist only of the deterministic timing model and Gaussian white noise with an amplitude of $10^{-7}$ seconds.\\
  
  \item[Simulation 2:] As simulation 1, however with a white noise amplitude of $10^{-6}$ seconds.\\
  
  \item[Simulation 3:] As simulation 1, however with the addition of a red noise signal described by Eq.\ref{Eq:Red} with $A=5\times10^{-14}$ and $\gamma=4.333$. \\

\end{description}

In all three simulations we use a simulated timing model for the binary pulsar J1713+0747 consistent with current observed values \citealt{2005ApJ...620..405S}, details of which are given in Table \ref{Table:pulsarparams}. When performing the linearised parameterisation of the timing model we perform the linearisation at the injected parameter values in order to maximise the performance of the method, and thus provide the most stringent comparison.

\begin{table*}
\caption{Injected timing model parameter values for PSR J1713+0747}
\begin{tabular}{ll}
\hline\hline
\multicolumn{2}{c}{Fit and data-set} \\
\hline
Pulsar name\dotfill & J1713+0747 \\ 
MJD range\dotfill & 50000.3---53002.0 \\ 
Number of TOAs\dotfill & 216 \\
\hline
\multicolumn{2}{c}{Measured Quantities} \\ 
\hline
Right ascension, $\alpha$\dotfill &  17:13:49.5325545 \\ 
Declination, $\delta$\dotfill & +07:47:37.49998 \\ 
Pulse frequency, $\nu$ (s$^{-1}$)\dotfill & 218.81184044143486131 \\ 
First derivative of pulse frequency, $\dot{\nu}$ (s$^{-2}$)\dotfill & $-$4.0839248109419448511$\times 10^{-16}$ \\ 
Dispersion measure, $DM$ (cm$^{-3}$pc)\dotfill & 15.9936 \\ 
Proper motion in right ascension, $\mu_{\alpha}$ (mas\,yr$^{-1}$)\dotfill & 4.91612625 \\ 
Proper motion in declination, $\mu_{\delta}$ (mas\,yr$^{-1}$)\dotfill & $-$3.9208688 \\ 
Parallax, $\pi$ (mas)\dotfill & 0.9359045 \\ 
Orbital period, $P_b$ (d)\dotfill & 67.8251296839 \\ 
Epoch of periastron, $T_0$ (MJD)\dotfill & 54303.63538774 \\ 
Projected semi-major axis of orbit, $x$ (lt-s)\dotfill & 32.34242233904 \\ 
Longitude of periastron, $\omega_0$ (deg)\dotfill & 176.20415671 \\ 
Orbital eccentricity, $e$\dotfill & 7.4940259711$\times 10^{-5}$ \\ 
First derivative of orbital period, $\dot{P_b}$\dotfill & 3.00166$\times 10^{-13}$ \\ 
Periastron advance, $\dot{\omega}$ (deg/yr)\dotfill & $-$3.6932$\times 10^{-5}$ \\ 
Companion mass, $M_c$ ($M_\odot$)\dotfill & 0.3112297 \\ 
Longitude of ascending node, $\Omega$ (degrees)\dotfill & 93.90581 \\ 
Orbital inclination angle, $i$ (degrees)\dotfill & 71.139153 \\ 
\hline
\multicolumn{2}{c}{Set Quantities} \\ 
\hline
Epoch of frequency determination (MJD)\dotfill & 54312 \\ 
Epoch of position determination (MJD)\dotfill & 54312 \\ 
Epoch of dispersion measure determination (MJD)\dotfill & 54312 \\ 
\hline
\multicolumn{2}{c}{Assumptions} \\
\hline
Clock correction procedure\dotfill & TT(TAI) \\
Solar system ephemeris model\dotfill & DE421 \\
Binary model\dotfill & T2 \\
Model version number\dotfill & 5.00 \\ 
\hline
\end{tabular}
\label{Table:pulsarparams}
\end{table*}

\begin{figure*}
\captionsetup{width=0.5\textwidth}
\begin{minipage}{180mm}
\begin{center}$
\begin{array}{c}
\includegraphics[width=180mm,height=100mm]{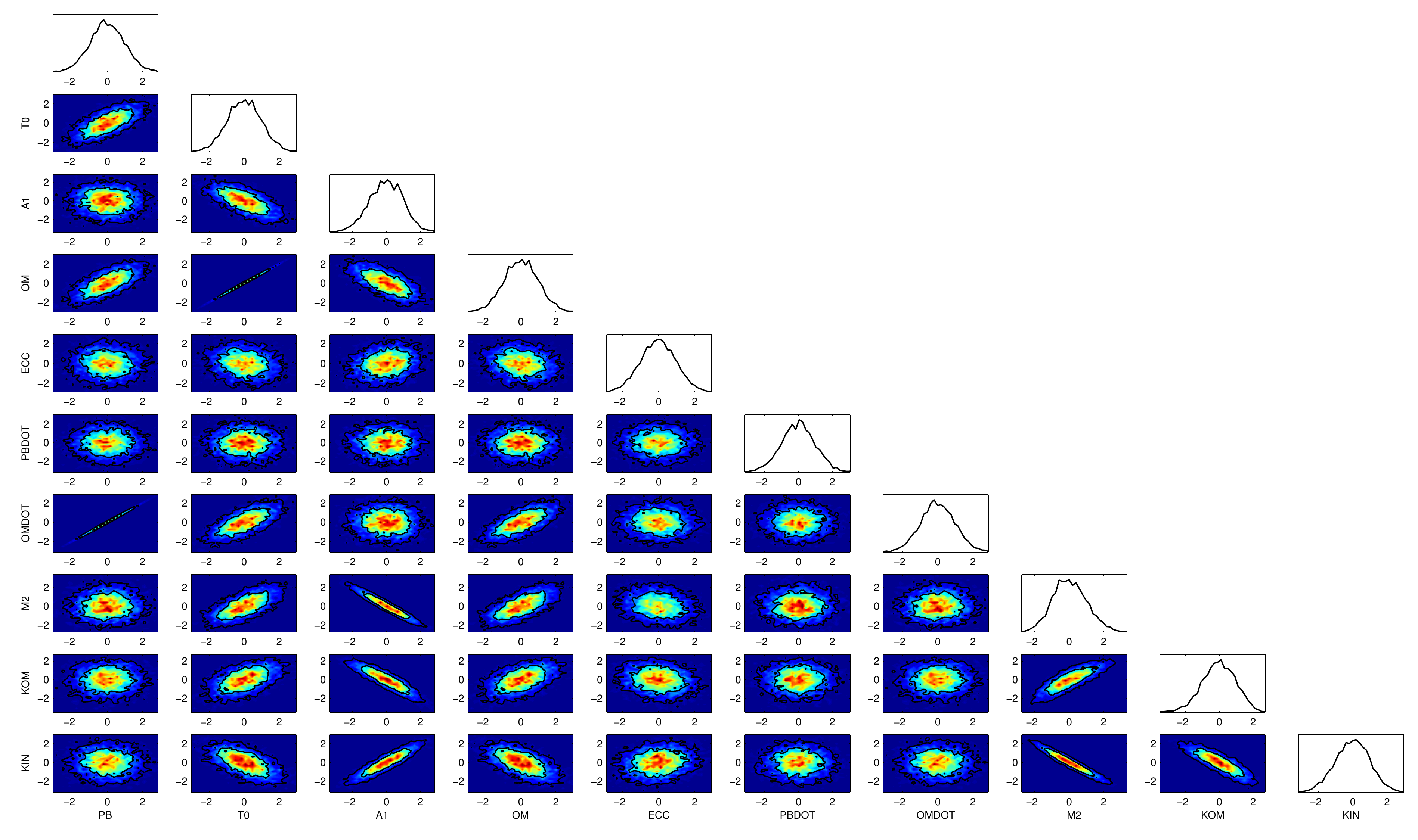} \\
\includegraphics[width=180mm,height=100mm]{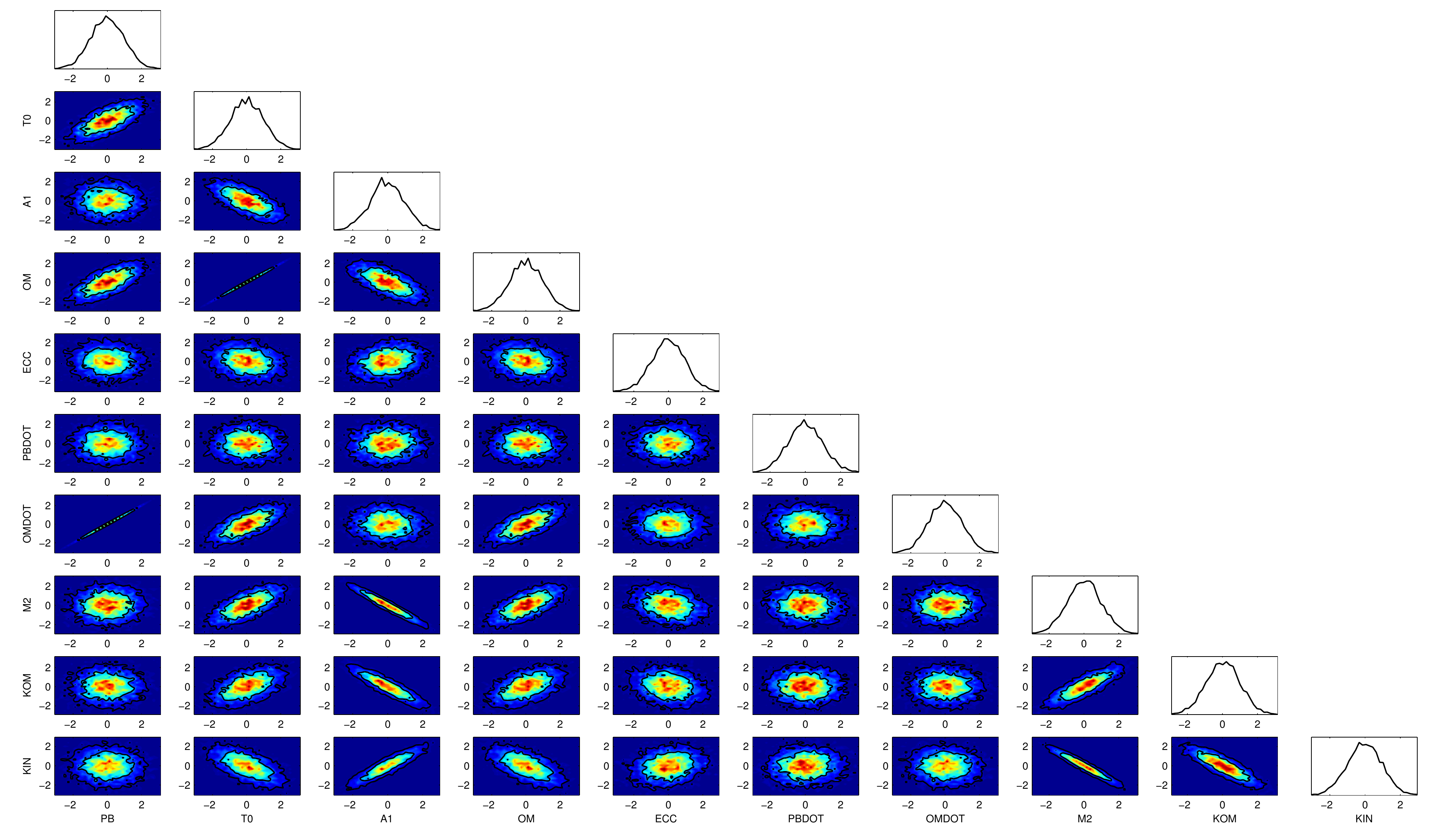}
\end{array}$
\end{center}
\caption{\protect One and two-dimensional marginalised posterior distributions for a subset of the binary parameters for PSR J1713+0747 for simulation 1 for the 
non-linear (top) and linear (bottom) timing models.  These parameters are from left to right: the orbital period of the binary (PB), the epoch of periastron (T0) 
the projected semi-major axis of orbit (A1), the longitude of periastron (OM), the eccentricity (ECC), the first derivative of the orbital period of the binary (PBDOT), 
the first derivative of the longitude of periastron (OMDOT), the companion mass (M2), the longitude of ascending node (KOM) and the inclination angle (KIN). 
 In all cases the scale on the x-axis is the deviation from the injected parameter values in units of the uncertainty in the parameter returned by Tempo2. 
  In the high signal--to--noise regime of these simulations the two models are completely consistent with one another, both in terms of parameter estimates, 
  and uncertainties.}
\label{figure:Sim1}
\end{minipage}
\end{figure*}

\begin{figure*}
\begin{minipage}{180mm}
\begin{center}$
\begin{array}{c}
\includegraphics[width=180mm,height=100mm]{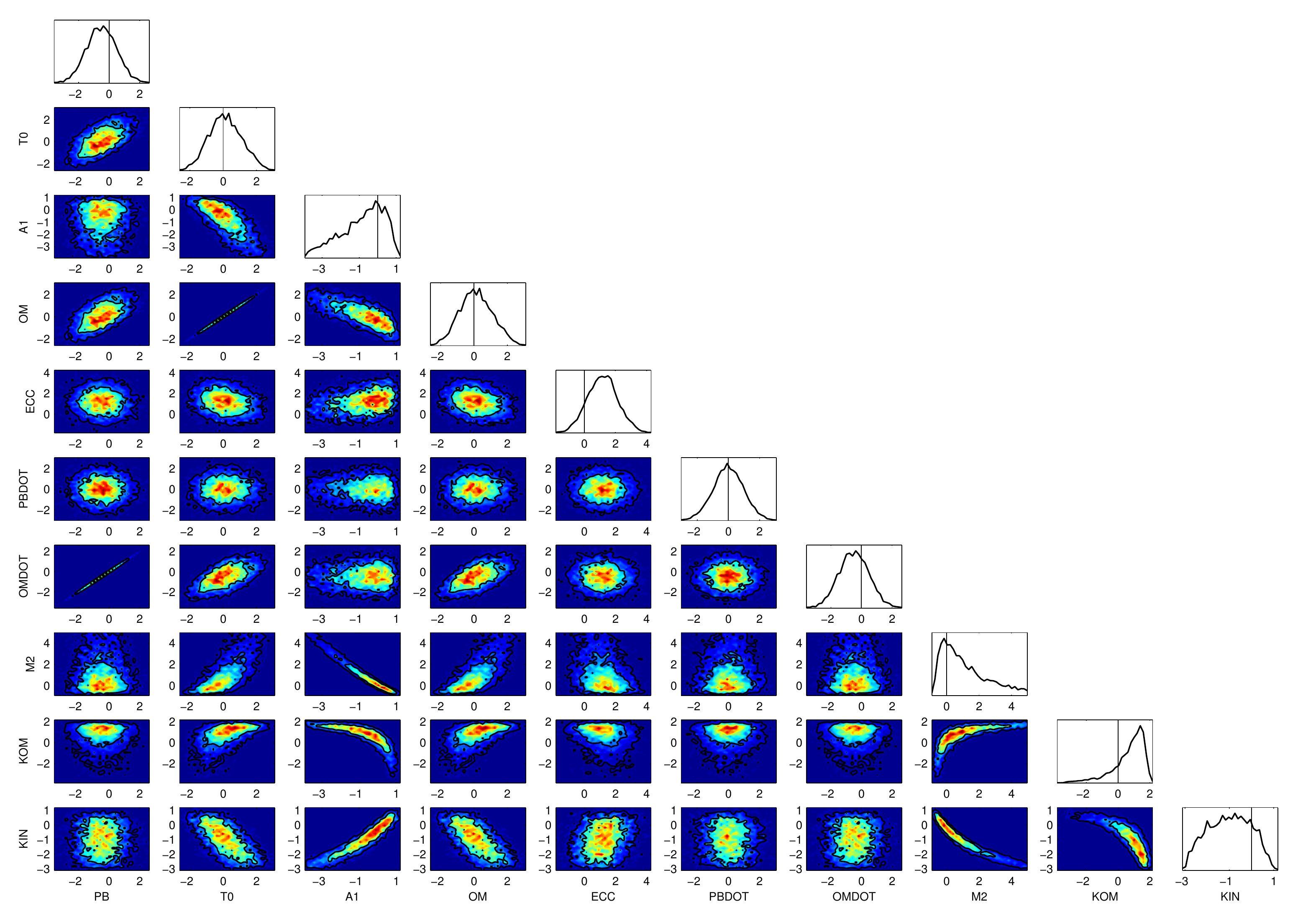} \\
\includegraphics[width=180mm,height=100mm]{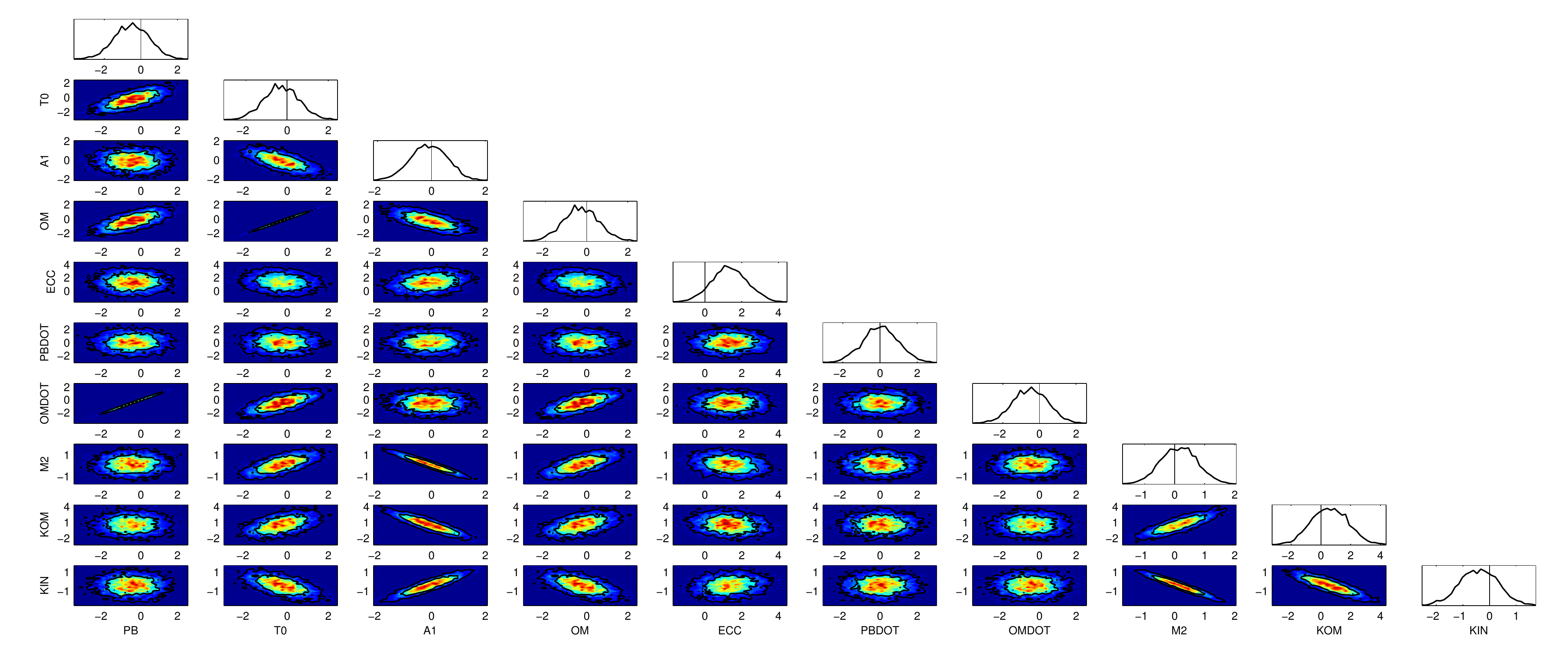}
\end{array}$
\end{center}
\caption{One and two-dimensional marginalised posterior distributions for a subset of the binary parameters for PSR J1713+0747 for simulation 2 for the  
non-linear (top) and linear (bottom) timing models.  These parameters are from left to right: the orbital period of the binary (PB), the epoch of periastron (T0)  
the projected semi-major axis of orbit (A1), the longitude of periastron (OM), the eccentricity (ECC), the first derivative of the orbital period of the binary (PBDOT),  
the first derivative of the longitude of periastron (OMDOT), the companion mass (M2), the longitude of ascending node (KOM) and the inclination angle (KIN).  
In all cases the scale on the x-axis is the deviation from the injected parameter values in units of the uncertainty in the parameter returned by Tempo2.  With the  
increase in the level of the white noise ($1\mu s$) there are now significant differences in the posterior distributions of the two timing models with large non-Gaussian  
 tails leading to an under-estimation of the uncertainties in the linear model.}
\label{figure:Sim2}
\end{minipage}
\end{figure*}

\begin{figure*}
\begin{minipage}{180mm}
\begin{center}$
\begin{array}{c}
\includegraphics[width=180mm,height=100mm]{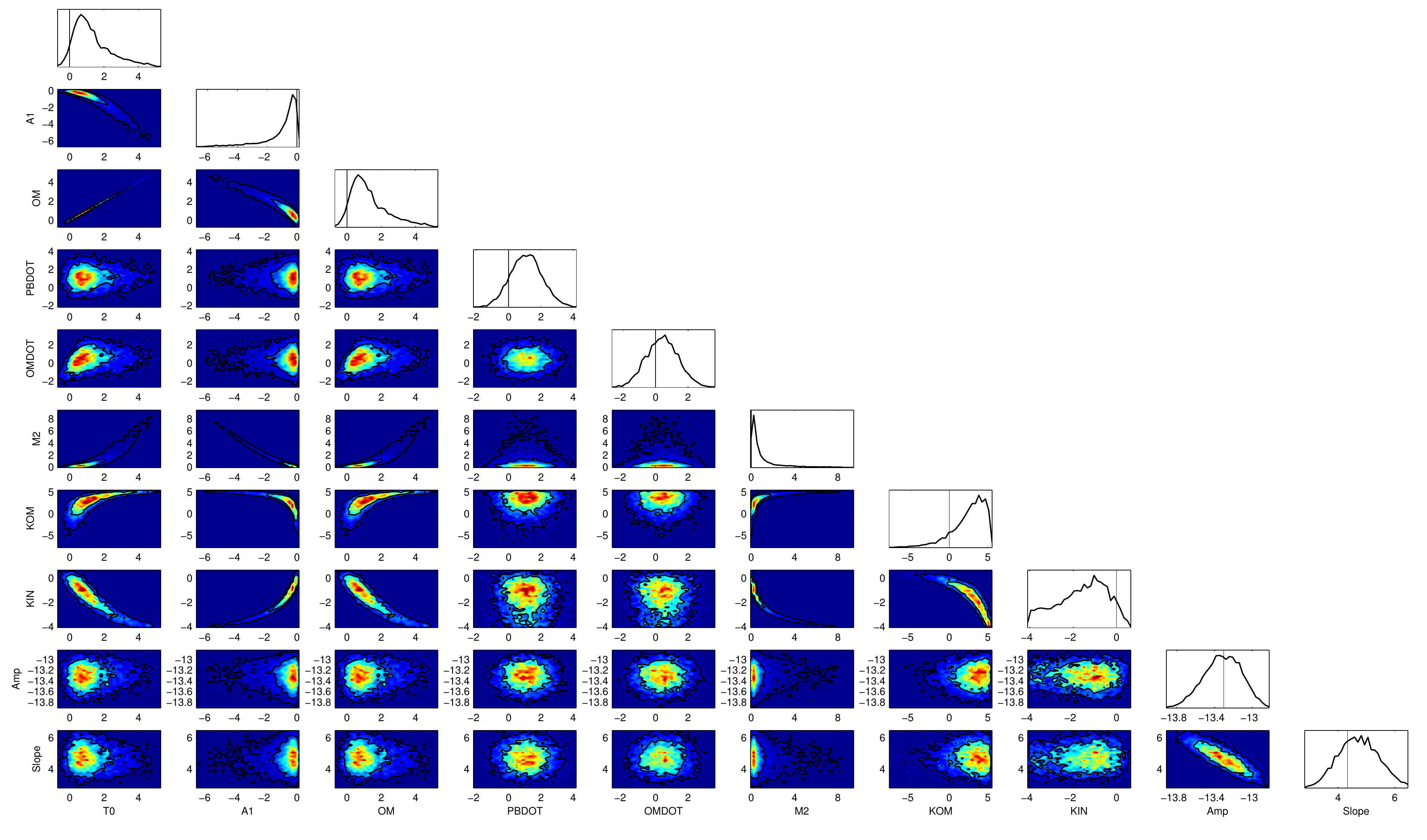} \\
\includegraphics[width=180mm,height=100mm]{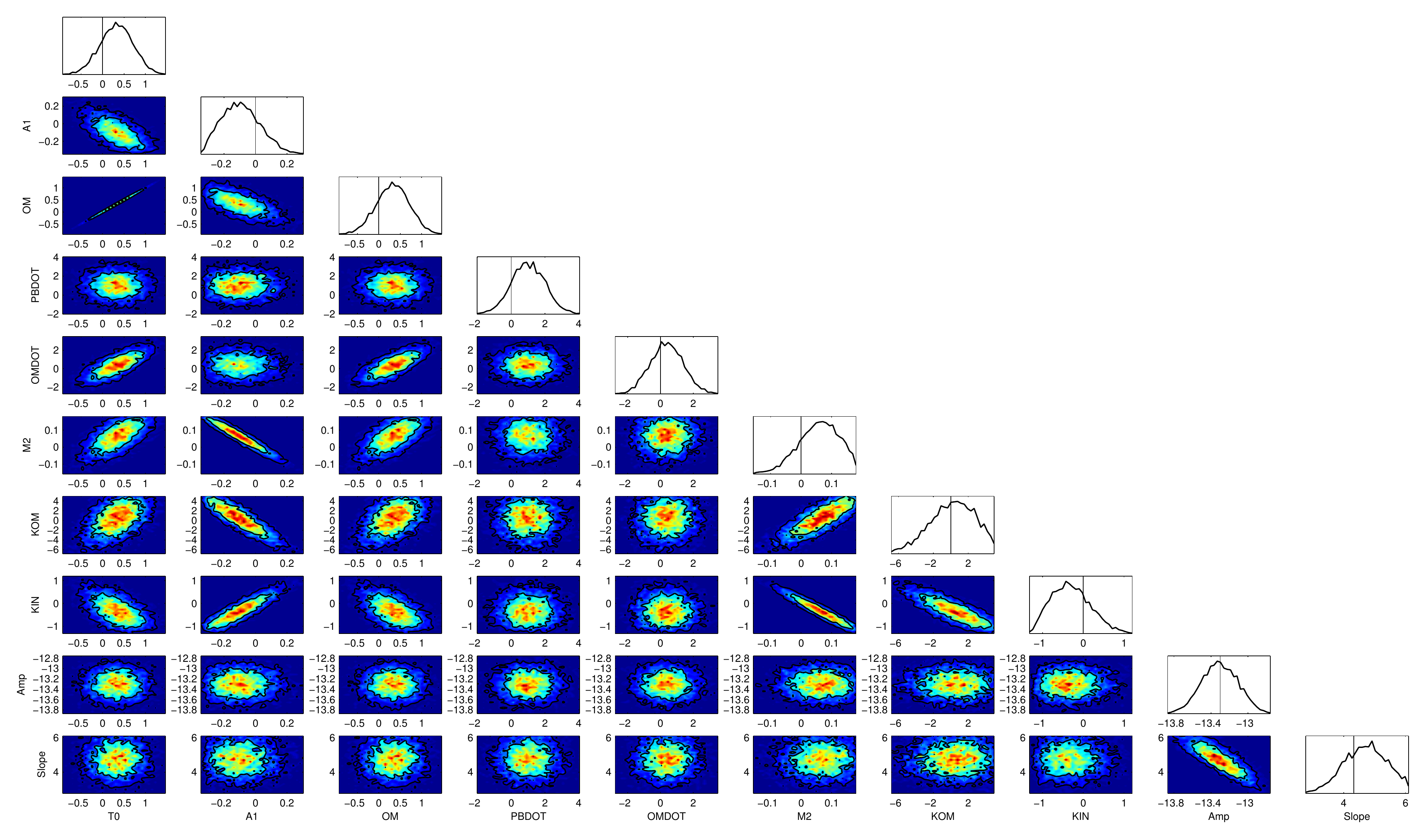}
\end{array}$
\end{center}
\caption{One and two-dimensional marginalised posterior distributions for a subset of the binary parameters for PSR J1713+0747 for simulation 3 for the  
 non-linear (top) and linear (bottom) timing models.  These parameters are from left to right: the epoch of periastron (T0), the projected semi-major axis of orbit (A1),  
 the longitude of periastron (OM), the first derivative of the orbital period of the binary (PBDOT), the first derivative of the longitude of periastron (OMDOT), the companion  
 mass (M2), the longitude of ascending node (KOM) and the inclination angle (KIN).  In all cases but the red noise parameters, the scale on the x-axis is the deviation from 
 the injected parameter values in units of the uncertainty in the parameter returned by Tempo2. Whilst all parameters are consistent with the injected values within 2$\sigma$
  confidence internals, the addition of red noise to the signal has resulted in even greater discrepancies in the estimated parameter uncertainties between the linear and  
  non-linear timing models, however the posterior distributions for the two stochastic parameters are extremely consistent between both.}
\label{figure:Sim3}
\end{minipage}
\end{figure*}

Tables \ref{Table:Sim1} to \ref{Table:Sim3} show the maximum likelihood, linear and non-linear timing model and stochastic parameter estimates for the three simulations.  In all cases we list only a single set of maximum likelihood parameter estimates, as these are the same for both linear and non-linear models.  Figs. \ref{figure:Sim1} to \ref{figure:Sim3} then show the one and two dimensional marginalised posteriors for a subset of the non-linear (top plots) and linear (bottom plots) timing model parameters related to the binary properties of the system that show the greatest differences when comparing the two models.  For simulation 3 we substitute two of the timing model parameters in these plots (the orbital period and eccentricity) in favour of the spectral index and amplitude of the red noise.

For the three simulations all the posterior distributions for the timing model parameters shown in Figs. \ref{figure:Sim1} to \ref{figure:Sim3} are consistent with the injected parameter values within 2$\sigma$ confidence intervals for both the linear and non-linear timing models.
From simulation 1 we see that in the high signal--to--noise regime, there is almost no observable difference between either the parameter estimates or their uncertainties for the linear and non-linear timing models.  This is to be expected as the range of parameter space over which the likelihood remains high is small, and thus the linear approximation should be valid.  

As we increase the level of the white noise however, from an amplitude of 100ns to $1\mu s$, we begin to see some significant differences between the two models.  In particular the companion mass and Kopeikin parameters (KOM and KIN) show large curving degeneracies between the parameters.  These non-Gaussian features are lost when we transition to the linear regime, which has the result of incorrectly estimating the uncertainties in these parameters.  For example, the 1$\sigma$ confidence intervals for the companion mass M2 is a factor 2.4 times smaller in the linear regime when compared to the non-linear.

This effect is accentuated even further in simulation three where we introduce a red noise signal into the data.  Here almost all parameters shown in Fig. \ref{figure:Sim3} show an underestimation of the error in the linear regime, with the most extreme examples showing 1$\sigma$ confidence intervals 8.8 and 10.8 times greater in the non-linear model for parameters $A1$ and $M2$ respectively.  It is important to note that in both cases we are modelling the red noise in the same way, and thus this effect is solely due to the differences between the linear and non-linear timing models.  As such, any method of pulsar timing analysis that operates in the linear regime, regardless of how it incorporates additional stochastic processes, such as the Cholesky method in C11, will suffer from this effect.

\begin{figure*}
\begin{minipage}{180mm}
\begin{center}$
\begin{array}{c}
\includegraphics[width=150mm]{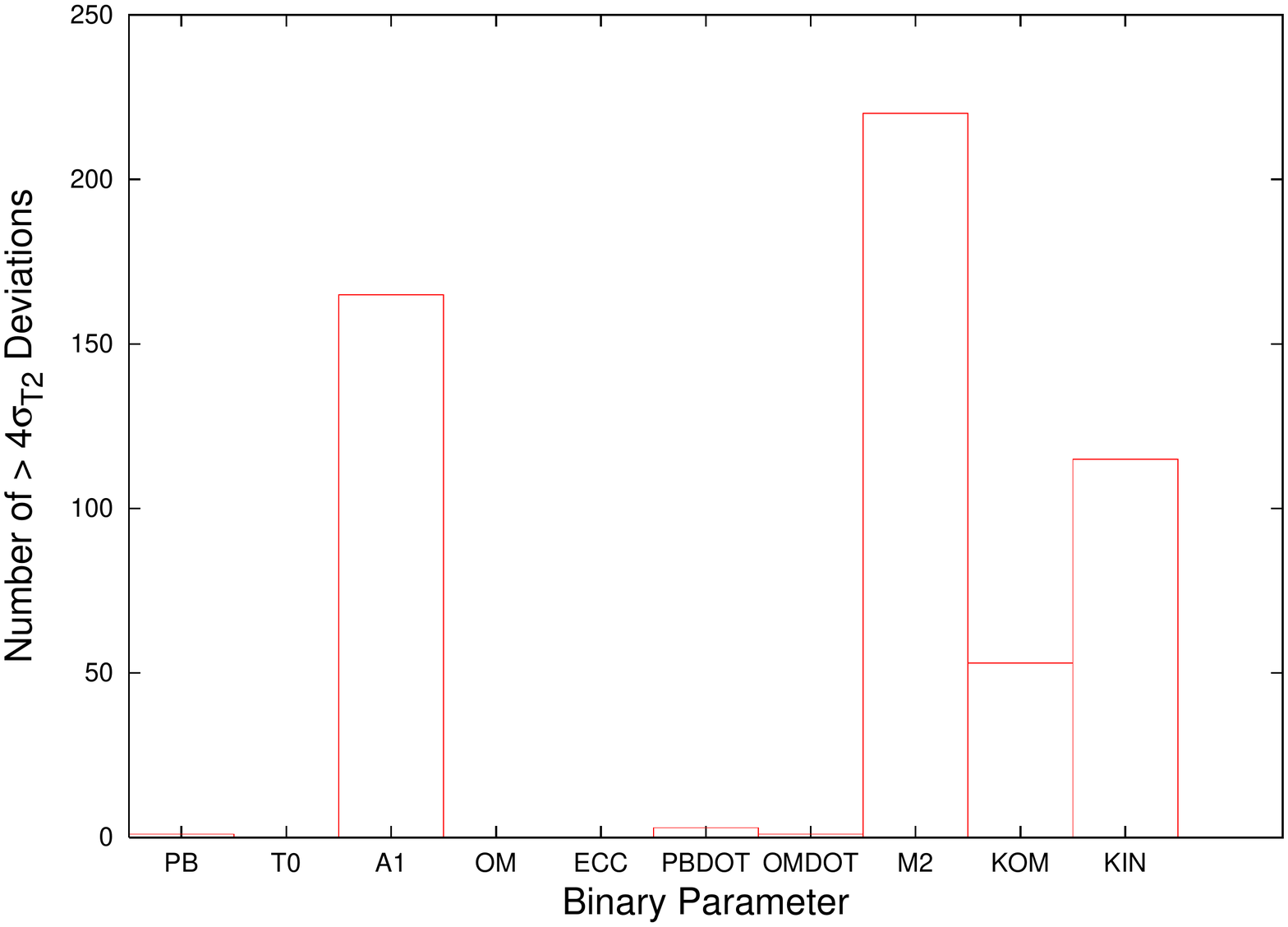} \\
\end{array}$
\end{center}
\caption{Number of greater than 4$\sigma_{T2}$ deviations for the binary parameters in the timing model of PSR J1713+0747 between the fit returned by Tempo2 using the linear timing model and the true value in a series of 10591 realisations of the noise in simulation 2, with $\sigma_{T2}$ the 1$\sigma$ uncertainty returned by Tempo2.  Given Gaussian statistics we would expect $\sim10$ such events across all parameters total.  The model parameters A1, M2, KIN and KOM show significant deviations from this prediction, indicating that the linear timing model significantly underestimates the errors in these parameters.  Comparing this result to Fig. \ref{figure:Sim2} we see that these parameters correspond to those that have large curving degeneracies in the posterior probability distribution, with significantly larger 1$\sigma$ confidence intervals than those returned by the linear model. }
\label{figure:sigmadev}
\end{minipage}
\end{figure*}

In this one set of simulations we do not see any significant bias in the timing model parameters returned by the linear timing model, however, given that the posterior probability distribution of the non-linear timing model shows significant curving degeneracies, and much greater $1\sigma$ confidence intervals for the binary parameters when compared to the linear model as the noise increases, we would expect that over a large number of realisations the number of occurences of $>4\sigma$ deviations that occur in the linear timing model should exceed that predicted by Gaussian statistics.  
In order to test this hypothesis we generate a series of 10591 realisations of the noise in simulation 2, using Tempo2 to calculate the parameter estimates with the linear timing model, and count the number of $>4\sigma_{T2}$ deviations for the binary parameters in the timing model, with $\sigma_{T2}$ the $1\sigma$ uncertainty returned by Tempo2.  Given Gaussian statistics we would expect $\sim10$ such events across all parameters total.  Fig. \ref{figure:sigmadev} shows a histogram for the number of events for the binary parameters in PSR J1713+0747. The model parameters A1, M2, KIN and KOM show a significant excess from the Gaussian prediction, indicating that the linear timing model significantly underestimates the errors in these parameters.  Comparing this result to Fig. \ref{figure:Sim2} we see that these parameters correspond to those that have large curving degeneracies in the posterior probability distribution, with significantly larger 1$\sigma$ confidence intervals than those returned by the linear model, confirming our hypothesis.

\begin{table*}
\centering
\caption{Maximum likelihood, non-Linear and linear timing model parameter estimates for Simulation 1}
\begin{tabular}{lccc}
\hline\hline
& Maximum Likelihood & Non-linear & Linear \\
\hline
\hline
Right ascension, $\alpha$\dotfill &  4.510914902 & 4.510914902 (3) &  4.510914902 (3)\\ 
Declination, $\delta$\dotfill & 0.1360265985 & 0.1360265984 (3) & 0.1360265984 (3)\\ 
Pulse frequency, $\nu$ (s$^{-1}$)\dotfill & 218.81184044143492 & 218.81184044143486 (18) &218.81184044143486 (18)\\ 
First derivative of pulse frequency, $\dot{\nu}$ (s$^{-2}$)\dotfill & $-$4.083922 $\times 10^{-16}$ & $-$4.083925 (6) $\times 10^{-16}$ &  $-$4.083922 (6) $\times 10^{-16}$\\ 
Proper motion in right ascension, $\mu_{\alpha}$ (mas\,yr$^{-1}$)\dotfill & 4.9186 & 4.9161 (18) &  4.9161 (18)\\ 
Proper motion in declination, $\mu_{\delta}$ (mas\,yr$^{-1}$)\dotfill & $-$3.922  & $-$3.921 (3) & $-$3.921 (3)\\ 
Parallax, $\pi$ (mas)\dotfill & 0.94  & 0.94 (2)& 0.94 (2) \\ 
Orbital period, $P_b$ (d)\dotfill & 67.825130 & 67.825129 (7) &  67.825130 (7)\\ 
Epoch of periastron, $T_0$ (MJD)\dotfill & 54303.6354 & 54303.6354 (4) & 54303.6354 (4) \\ 
Projected semi-major axis of orbit, $x$ (lt-s)\dotfill & 32.34242227 & 32.34242233 (16) &  32.34242233 (16) \\ 
Longitude of periastron, $\omega_0$ (deg)\dotfill & 176.204  & 176.204 (2) & 176.204 (2)\\ 
Orbital eccentricity, $e$\dotfill & 7.49410$\times 10^{-5}$ & 7.49402 (6) $\times 10^{-5}$ &  7.49402 (6) $\times 10^{-5}$\\ 
First derivative of orbital period, $\dot{P_b}$\dotfill & 2$\times 10^{-14}$& 3 (6)$\times 10^{-13}$ & 3 (6)$\times 10^{-13}$\\ 
Periastron advance, $\dot{\omega}$ (deg/yr)\dotfill & $-$4$\times 10^{-5}$ & $-$3 (19) $\times 10^{-5}$ &$-$3 (19) $\times 10^{-5}$ \\ 
Companion mass, $M_c$ ($M_\odot$)\dotfill & 0.316 & 0.312 (16)&  0.312 (16)\\ 
Longitude of ascending node, $\Omega$ (degrees)\dotfill & 94.1  & 93.8 (1.4)&93.9 (1.4) \\ 
Orbital inclination angle, $i$ (degrees)\dotfill & 71.0 & 71.1 (6)&71.1 (7) \\ 
\hline
\end{tabular}
\label{Table:Sim1}
\end{table*}

\begin{table*}
\caption{Maximum likelihood, non-linear and linear timing model parameter estimates for Simulation 2}
\begin{tabular}{lccc}
\hline\hline
& Maximum Likelihood & Non-Linear & Linear \\
\hline
\hline
Right ascension, $\alpha$\dotfill &  4.51091490 & 4.51091486 (3) &  4.51091486 (2)\\ 
Declination, $\delta$\dotfill & 0.136026598 & 0.136026602 (3) & 0.136026602 (3)\\ 
Pulse frequency, $\nu$ (s$^{-1}$)\dotfill & 218.8118404414366 & 218.8118404414348 (18) & 218.8118404414348 (18)\\ 
First derivative of pulse frequency, $\dot{\nu}$ (s$^{-2}$)\dotfill & $-$4.08385 $\times 10^{-16}$ & $-$4.08392 (6) $\times 10^{-16}$ &  $-$4.08392 (6) $\times 10^{-16}$\\ 
Proper motion in right ascension, $\mu_{\alpha}$ (mas\,yr$^{-1}$)\dotfill & 4.906 & 4.905 (17) &  4.906 (17)\\ 
Proper motion in declination, $\mu_{\delta}$ (mas\,yr$^{-1}$)\dotfill & $-$3.90  & $-$3.91 (3) & $-$3.91 (3)\\ 
Parallax, $\pi$ (mas)\dotfill & 1.2  & 1.0 (2)& 0.9 (2) \\ 
Orbital period, $P_b$ (d)\dotfill & 67.82513 & 67.82510 (6) &  67.82510 (6)\\ 
Epoch of periastron, $T_0$ (MJD)\dotfill & 54303.635 & 54303.636 (4) & 54303.634 (4) \\ 
Projected semi-major axis of orbit, $x$ (lt-s)\dotfill & 32.342422 & 32.342421 (22) &  32.342422 (16) \\ 
Longitude of periastron, $\omega_0$ (deg)\dotfill & 176.20  & 176.21 (2) & 176.20 (2)\\ 
Orbital eccentricity, $e$\dotfill & 7.4948$\times 10^{-5}$ & 7.4948 (6) $\times 10^{-5}$ &  7.4949 (6) $\times 10^{-5}$\\ 
First derivative of orbital period, $\dot{P_b}$\dotfill & $-$1.7$\times 10^{-12}$& 1.5 (65)$\times 10^{-13}$ & 5 (62)$\times 10^{-13}$\\ 
Periastron advance, $\dot{\omega}$ (deg/yr)\dotfill & $-$4$\times 10^{-5}$ & $-$9 (18) $\times 10^{-4}$ &$-$10 (19) $\times 10^{-4}$ \\ 
Companion mass, $M_c$ ($M_\odot$)\dotfill & 0.3 & 0.53 (27)&  0.34 (13)\\ 
Longitude of ascending node, $\Omega$ (degrees)\dotfill & 95  & 103 (13) & 102 (16) \\ 
Orbital inclination angle, $i$ (degrees)\dotfill & 69 & 63 (8) & 68 (6) \\ 
\hline
\end{tabular}
\label{Table:Sim2}
\end{table*}

\begin{table*}
\caption{Maximum likelihood, non-linear and linear timing model parameter estimates for Simulation 3}
\begin{tabular}{lccc}
\hline\hline
& Maximum Likelihood & Non-Linear & Linear \\
\hline
\hline
Right ascension, $\alpha$\dotfill &  4.51091491  & 4.51091487 (2) &  4.510914907 (9)\\ 
Declination, $\delta$\dotfill &0.136026597  & 0.136026601  (3) & 0.1360265978 (18)\\ 
Pulse frequency, $\nu$ (s$^{-1}$)\dotfill & 218.8118404414 & 218.8118404412 (2) & 218.8118404412 (2) \\ 
First derivative of pulse frequency, $\dot{\nu}$ (s$^{-2}$)\dotfill & $-$4.084 $\times 10^{-16}$ & $-$4.091 (4) $\times 10^{-16}$  &  $-$4.090 (4) $\times 10^{-16}$\\ 
Proper motion in right ascension, $\mu_{\alpha}$ (mas\,yr$^{-1}$)\dotfill & 4.93 & 4.94(2) &  4.93 (2)\\ 
Proper motion in declination, $\mu_{\delta}$ (mas\,yr$^{-1}$)\dotfill & $-$3.94   & $-$3.93 (4) & $-$3.93 (4)\\ 
Parallax, $\pi$ (mas)\dotfill &0.5  & 0.6 (2)& 0.4 (2) \\ 
Orbital period, $P_b$ (d)\dotfill & 67.82513  & 67.82516 (7) &  67.82515 (7)\\ 
Epoch of periastron, $T_0$ (MJD)\dotfill &54303.635  & 54303.649 (11) & 54303.639 (4) \\ 
Projected semi-major axis of orbit, $x$ (lt-s)\dotfill & 32.342422  & 32.342412 (12) &  32.3424211 (12) \\ 
Longitude of periastron, $\omega_0$ (deg)\dotfill & 176.20    & 176.28 (6) & 176.22 (2)\\ 
Orbital eccentricity, $e$\dotfill & 7.4938$\times 10^{-5}$ & 7.4932 (8) $\times 10^{-5}$ &  7.4936 (6) $\times 10^{-5}$\\ 
First derivative of orbital period, $\dot{P_b}$\dotfill & 4$\times 10^{-12}$& 7 (6)$\times 10^{-12}$ & 6 (6)$\times 10^{-12}$\\ 
Periastron advance, $\dot{\omega}$ (deg/yr)\dotfill & 0.0016  & 0.0007 (18)  & 0.0007 (19) \\ 
Companion mass, $M_c$ ($M_\odot$)\dotfill & 0.3   & 2 (3)&  0.44 (13)\\ 
Longitude of ascending node, $\Omega$ (degrees)\dotfill & 91   & 108 (13) & 98 (18) \\ 
Orbital inclination angle, $i$ (degrees)\dotfill & 72  & 52 (13) & 66 (7) \\ 
$\log_{10} \mathrm{A_{red}}$\dotfill &-13.01 & $-$13.31 (19)& -13.31 (20) \\
Spectral Index \dotfill & 3.5  & 4.7(7) & 4.7 (7)\\
\hline
\end{tabular}
\label{Table:Sim3}
\end{table*}

\section{Application to real data}
\label{Section:RealData}

We now demonstrate the application of TempoNest to the publicly available datasets for the binary pulsar B1855+09 and the isolated pulsar B1937+21 presented in \citep{1994ApJ...428..713K} (henceforth K94).  For the former, we demonstrate the ability of TempoNest to perform rigorous model selection between different sets of stochastic and timing model parameters, whilst for the latter we compare power law and model independent descriptions of the red spin noise, and in addition to these also compare the method of \citep{2013MNRAS.429.2161K} in our analysis of the dispersion measure variations in order to find the optimal description of the stochastic properties of the data.

\subsection{B1855+09}

\begin{table*}
\caption{Timing model and stochastic parameter estimates for PSR B1855+09}
\begin{tabular}{lll}
\hline\hline
\multicolumn{3}{c}{Fit and data-set} \\
\hline
Pulsar name\dotfill & B1855+09 \\ 
MJD range\dotfill & 46436.7---48973.7 \\ 
Number of TOAs\dotfill & 270 \\
Evidence \dotfill & 3450.2 $\pm$ 0.3 \\
\hline
\multicolumn{3}{c}{Measured Quantities} \\ 
\hline
& TempoNest & Tempo2 \\
\hline
Right ascension, $\alpha$\dotfill & 18:57:36.394354(4) & 18:57:36.394354(4) \\ 
Declination, $\delta$\dotfill & +09:43:17.31966(10) & +09:43:17.31966(11)\\ 
Pulse frequency, $\nu$ (s$^{-1}$)\dotfill & 186.49440787786523(4) & 186.49440787786523(4)\\ 
First derivative of pulse frequency, $\dot{\nu}$ (s$^{-2}$)\dotfill & $-$6.20499(9)$\times 10^{-16}$ & $-$6.20500(9)$\times 10^{-16}$\\ 
Dispersion measure, $DM$ (cm$^{-3}$pc)\dotfill & 13.307(3) & 13.308(3)\\ 
Proper motion in right ascension, $\mu_{\alpha}$ (mas\,yr$^{-1}$)\dotfill & $-$2.63(3)& $-$2.63(3) \\ 
Proper motion in declination, $\mu_{\delta}$ (mas\,yr$^{-1}$)\dotfill & $-$5.41(5) &  $-$5.46(5)\\ 
Parallax, $\pi$ (mas)\dotfill & 1.2(2) & 1.1(3)\\ 
Sine of inclination angle,$\sin{i}$\dotfill & 0.9991(4)& 0.9990(4) \\ 
Orbital period, $P_b$ (d)\dotfill & 12.3271713813(4)& 12.3271713815(5) \\ 
Epoch of periastron, $T_0$ (MJD)\dotfill & 47529.896(2) & 47529.8966(19)\\ 
Projected semi-major axis of orbit, $x$ (lt-s)\dotfill & 9.2307801(3) & 9.2307802(3) \\ 
Longitude of periastron, $\omega_0$ (deg)\dotfill & 276.39(6) & 276.39(6)\\ 
Orbital eccentricity, $e$\dotfill & 2.170(3)$\times 10^{-5}$ & 2.169(4)$\times 10^{-5}$ \\ 
Companion mass, $M_c$ ($M_\odot$)\dotfill & 0.270(14)& 0.265(14)\\ 
\hline
\multicolumn{3}{c}{Stochastic Parameters} \\
\hline
EFAC \dotfill & 0.806(11) & - \\
$\log_{10}$[EQUAD] \dotfill & -6.2(2) & - \\
$\log_{10}$ [RedC1] \dotfill& -4.5(1.0)& - \\ 
\hline
\multicolumn{3}{c}{Set Quantities} \\ 
\hline
Epoch of frequency determination (MJD)\dotfill & 47526 \\ 
Epoch of position determination (MJD)\dotfill & 47526 \\ 
Epoch of dispersion measure determination (MJD)\dotfill & 47526 \\ 
\hline
\multicolumn{3}{c}{Assumptions} \\
\hline
Clock correction procedure\dotfill & TT(TAI) \\
Solar system ephemeris model\dotfill & DE405 \\
Binary model\dotfill & T2 \\
Model version number\dotfill & 5.00 \\ 
\hline
\multicolumn{3}{c}{Additional Included Parameters} \\
\hline
First derivitive of orbital eccentricity, $\dot{e}$\dotfill &  -2(5)$\times 10^{-16}$ & \\
First derivitive of orbital period, $\dot{P_b}$\dotfill & 0.2 (1.1)$\times 10^{-12}$ & \\
First derivitive of $x$, $\dot{x}$\dotfill & 1.5 (2.1)$\times 10^{-15}$ & \\
Periastron advance, $\dot{\omega}$ (deg/yr) \dotfill & -0.01(4) & \\
Jump 1 mk3\_14w \dotfill & -1.8 (1.1) $\times 10^{-6}$ & \\
Jump 2 mk3\_14m \dotfill & 0.5 (1.9) $\times 10^{-7}$ & \\
\hline
\end{tabular}
\label{Table:B1855}
\end{table*}

The mean posterior values and associated one-sigma errors for the final fitted timing model and stochastic parameters for PSR B1855+09 are listed in Table ~\ref{Table:B1855} and includes five astrometric quantities ($\alpha, \delta, \mu_{\alpha}, \mu_{\delta}, \pi$), two rotational parameters ($\nu, \dot{\nu}$), dispersion measure, as well as 7 binary parameters.  In addition to these we have included three stochastic parameters, an EFAC, EQUAD and a single red noise power spectrum coefficient, with a frequency equal to 1/$T$, with $T$ the total time span of the data.  

In performing the analysis using TempoNest we first performed a series of ten iterations with Tempo2 to ensure the timing solution had converged and set a uniform prior on the timing model parameters covering a range of $\pm10\sigma_{T2}$ from the maximum likelihood estimate obtained from the final iteration with $\sigma_{T2}$ the error returned by Tempo2.  The maximum likelihood Tempo2 estimates and one sigma errors are given in  Table ~\ref{Table:B1855} alongside the TempoNest results.  For the stochastic parameters we took our priors to be uniform across the ranges [0, 5], [$-10$, $-5$], [$-20$, 0] for EFAC, $\log_{10}\mathrm{EQUAD}$ and $\rho_i$ respectively.

In addition to these quantities, Table ~\ref{Table:B1855} lists the parameter estimates for 4 additional timing model parameters, $\dot{\omega}, \dot{P_b}, \dot{e}$ and $\dot{x}$ which were added to the timing model one at a time and the full analysis repeated.  In all cases however, the addition of the extra timing model parameters resulted in a decrease of the log-evidence by $\sim 1$ unit relative to the original fit indicating that there is no support for the parameters in the data.

Comparing the evidence for a model without the three stochastic parameters to that in Table ~\ref{Table:B1855} we find a decrease of the log-evidence of $\sim 2.5$ units.   Whilst this is not definitive, the inclusion of the stochastic parameters is still strongly favoured, and allows us to quantify the qualitative observation of a cubic signal present in the residuals described in \citep{1995IAUS..166..163K}.

Finally, because the observations of B1855+09 presented in K94 were made using 3 observing back-ends over the course of the dataset,  we also performed the analysis including two jumps between the different systems.  As there is a strong covariance between red noise signals and the jump parameters, we included the following combinations of parameters in our analysis :

\begin{description}
\item[Model 1:] Including Jumps - without any additional stochastic parameters
\item[Model 2:] Including Jumps - including EFAC/EQUAD
\item[Model 3:] Including Jumps - including EFAC/EQUAD and a single red noise coefficient at frequency 1/$T$.
\end{description}
As with the other timing model parameters, we set our prior to be $\pm10\sigma_{T2}$ from the Tempo2 initial estimate.  We could then compare the evidence returned from these analyses, with those models that exclude the jump parameters to see which set is most supported by the data.  In every instance the inclusion of the jumps resulted in either a small drop in the evidence of $\sim 0.5$, or it remained the same, suggesting no support for these parameters in the data.  The parameter estimates for the jumps when fitted alongside the optimal set of parameters are listed in Table ~\ref{Table:B1855} alongside the other additional parameters tested.

Comparing the parameter estimates obtained by TempoNest with those from Tempo2 we see that they are completely consistent for all values and their uncertainties.  Such agreement is unsurprising as the approximation that the data is well described by only the timing model and white noise is well justified in this instance given the minor support for additional stochastic parameters in the data.

\subsection{B1937+21}
\label{Section:1937}

In comparison to B1855+09 the timing model for pulsar B1937+21 is relatively simple, requiring only 8 parameters, the same 5 astrometric and 2 rotational quantities as for B1855+09, and dispersion measure.  However, the analysis is made more complex by the presence of significant long term variation in the timing residuals.  In order to account for one source of this noise, the TOAs in the K94 dataset were observed at two widely spaced frequencies, 1408 and 2380~MHz in order to calculate the effects of dispersion on the residuals prior to fitting the timing model.

In performing the analysis with TempoNest we therefore use two versions of the TOAs.  The first includes the DM corrections calculated in K94 (henceforth dataset 1), and the second excludes them (henceforth dataset 2).   The simple timing model required for this pulsar means that we expect little non-linearity despite the large amounts of noise present in the data.  We therefore first  analytically marginalise over the timing model using the timing model estimates obtained from Tempo2 and perform model selection between different sets of stochastic parameters for each dataset.  In each case we include an EFAC and EQUAD parameter, and then test different combinations of models for the red noise and DM variations in the data.   The full set of models compared in both datasets 1 and 2 are listed below.

\begin{itemize}
\item Dataset 1

\begin{description}
\item[Model 1:] model independent analysis with $N_c$ consecutive frequency coefficients
\item[Model 2:] model independent analysis with optimally chosen frequency coefficients
\item[Model 3:] power law analysis with $N_c$ consecutive coefficients
\end{description}

\item Dataset 2
\begin{description}
\item[Model 1:] model independent analysis with $N_c$ coefficients for both red noise and DM variations
\item[Model 2:] power law analysis with $N_c$ coefficients for red noise, model independent analysis of DM variations
\item[Model 3:] model independent analysis with $N_c$ coefficients for red noise, power law analysis for DM variations
\item[Model 4:] power law analysis with $N_c$ coefficients for both red noise and DM variations
\end{description}
\end{itemize}

\subsubsection{Dataset 1}
\label{Section:1939D1}

\begin{table}
\centering
\caption{Evidence for different stochastic models for pulsar B1937+21 in dataset 1} 
\centering 
\begin{tabular}{c c c} 
\hline\hline 
Model & $N_c$ & $\log$ Evidence\\[0.5ex] 
\hline
\hline 
model independent analysis & 5 & 0 \\
- & 6 & 2.8 \\
- & 7 & 2.7 \\
- & 8 & 2.7 \\
- & 10 & 2.3 \\
- & 15 & -2.1 \\
- & 20 & -6.5 \\
model independent analysis with optimal frequencies & 7 & 5.9 \\
power law  & 6 &  5.2\\
- & 10 &  6.5\\
- & 20 &  7.2\\
- & 50 &  7.4\\
- & 100 &  7.5\\
two component power law & 100 &  8.7\\
\hline
\end{tabular}
\label{Table:1937Ev} 
\end{table}

\begin{figure*}
\begin{minipage}{180mm}
\begin{center}$
\begin{array}{cc}
\includegraphics[width=85mm]{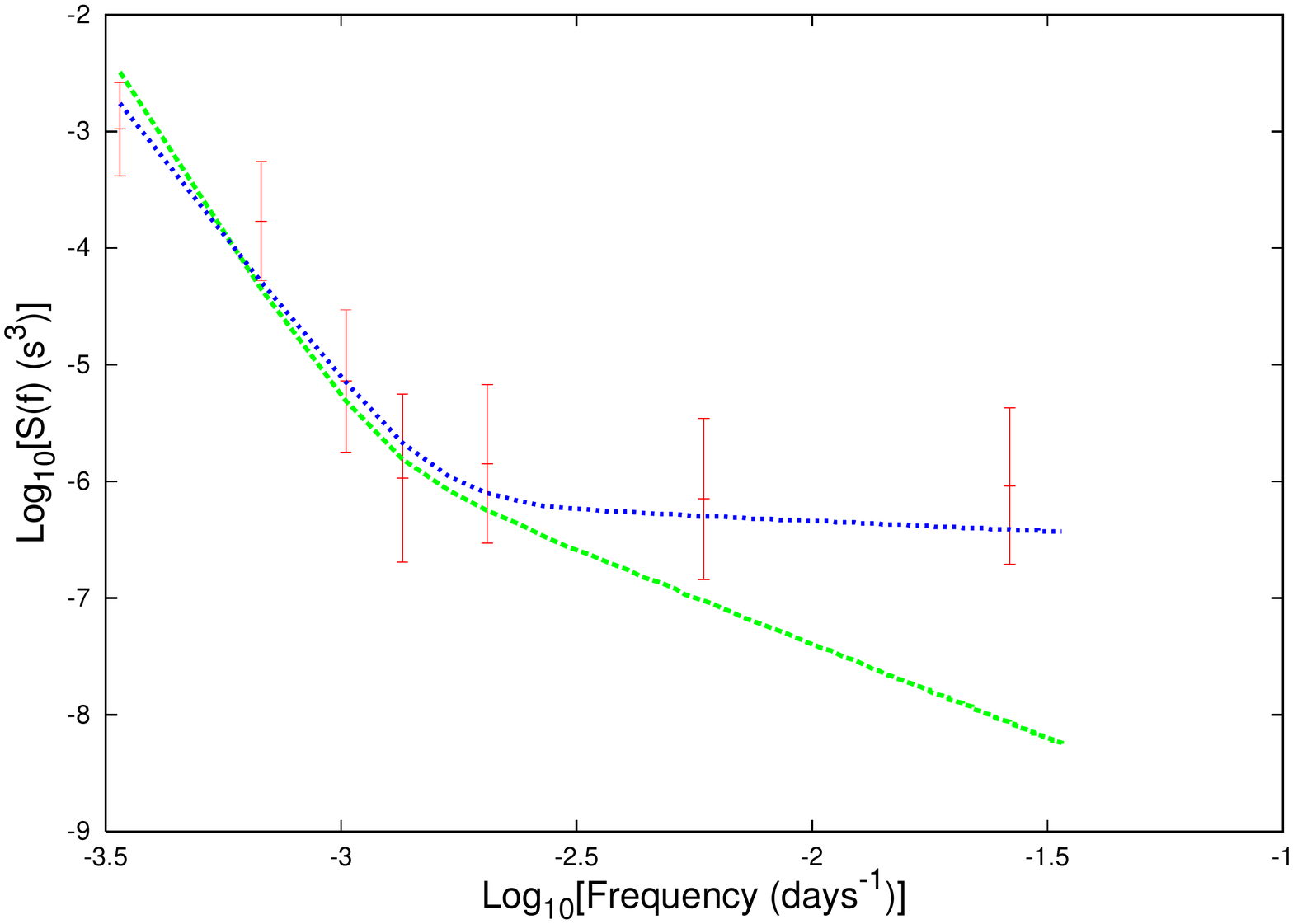} &
\includegraphics[width=85mm]{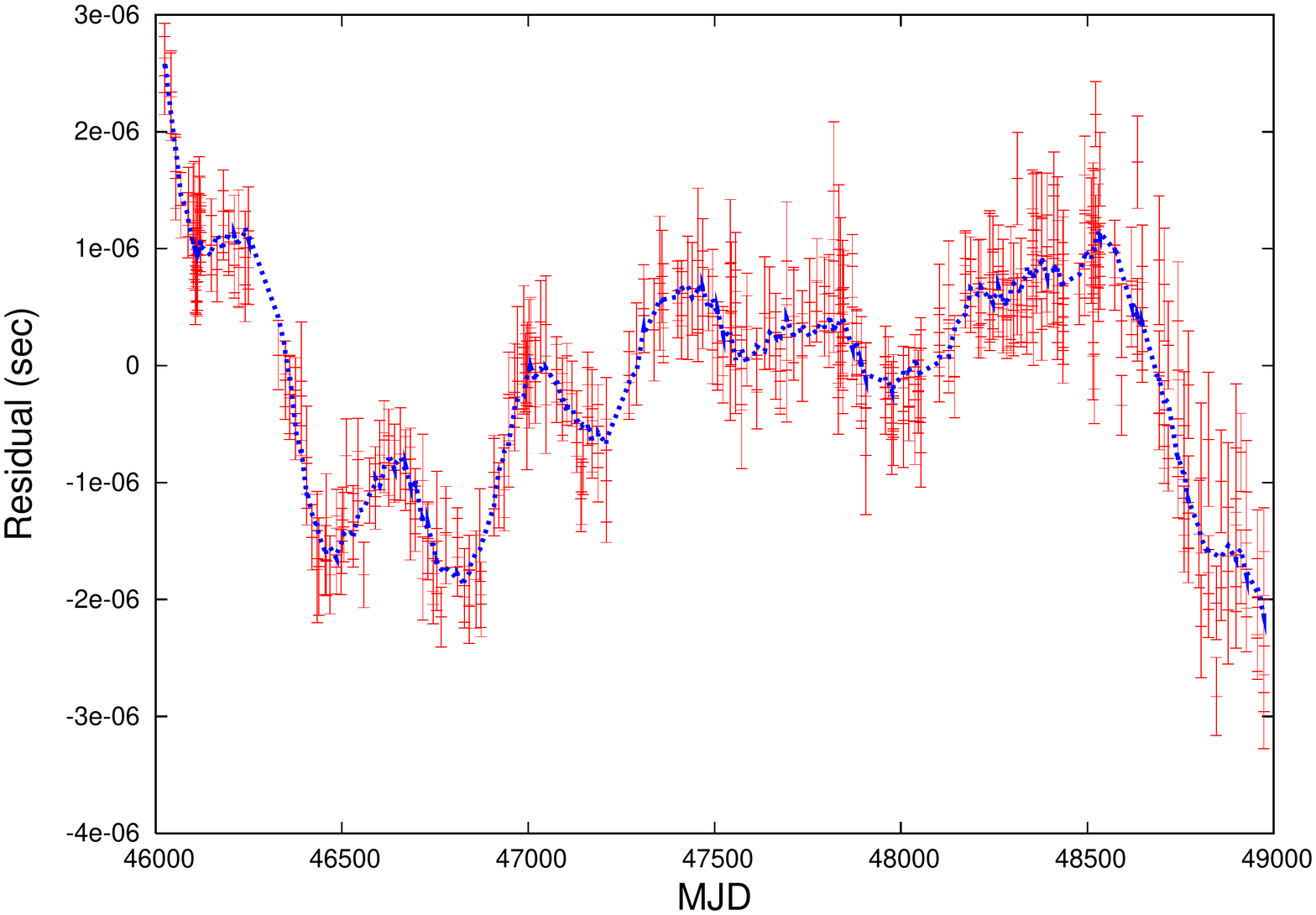} \\
\end{array}$
\end{center}
\caption{(left) The mean values and associated one sigma error bars for the power at the 7 
 optimal frequencies fitted to B1937+21 dataset 1 using the model independent analysis. The blue dotted and green dashed lines indicate the mean two component power law models fitted over the optimal set of frequencies only and over the first 100 consecutive frequencies respectively.   Whilst the two component model fitted to only the optimal set is consistent at all frequencies, we include this only for completion as we do not consider it to be a physically motivated model. (right) Timing residuals for B1937+21 dataset 1 (red points) and the best-fit signal realisation for the red noise using the optimal model independent analysis (blue dotted line).}
\label{figure:D1Res}
\end{minipage}
\end{figure*}

Table \ref{Table:1937Ev} shows the evidence returned for the different stochastic models applied to dataset 1.  Using the model independent description of the red noise signal we find that only 6 power spectrum coefficients are supported by the data when including consecutive frequencies.  Whilst this may seem like a small number for an apparently complex signal, in L13 it is shown that even in the high signal to noise regime, over an order of magnitude fewer power spectrum coefficients than time series data points are required to describe the data when dealing with steep red power spectrum.  

It is possible, however, that frequencies with $n>6$ are supported by the data but that considering only a consecutive set biases the model to include only low frequency coefficients.  To ascertain whether this is the case we perform the following test:

\begin{description}
\item[1:] Include the lowest 6 power spectrum coefficients in the model red noise model.
\item[2:] In addition include a coefficient with frequency a free parameter, allowed to vary continuously from $\nu = 6/T$ to $\nu = 100/T$.
\item[3:] Include all frequencies at which there is a peak in the posterior probability distribution for this floating coefficient into the model.
\item[4:] Eliminate coefficients until the optimal set is found, such that the Evidence is maximised.
\end{description}
A different approach to follow could be that of \cite{2011MNRAS.415.3462F}.  Here, the data would initially be analysed with one power spectrum coefficient, with frequency allowed to vary.  In this stage the evidence is not calculated, but the computationally less expensive process of parameter estimation is performed.  The resulting best fit model is subtracted from the data and a set of residuals formed.  An evidence calculation is then performed for two competing models on the residuals: i) That the residuals contain a signal described by one power spectrum coefficient, or ii) The data contains no signal.  If the evidence supports the inclusion of an additional power spectrum coefficient the parameter estimation is repeated with two components and a new set of residuals formed.  This process is then repeated until the evidence from the residual analysis no longer supports any signal.  This has the advantage that the evidence calculation need only be performed for a single frequency, eliminating much of the computational cost, however comparisons between this and other methods will be the subject of future work. 

\begin{figure}
\begin{minipage}{75mm}
\begin{center}$
\begin{array}{c}
\includegraphics[width=75mm]{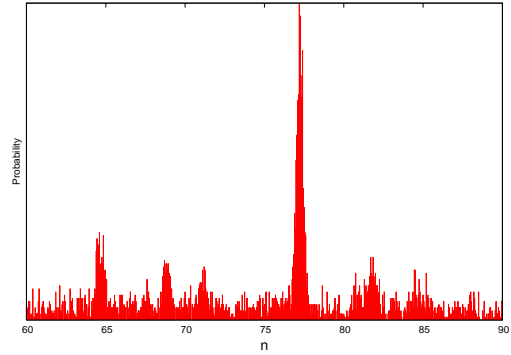} \\
\end{array}$
\end{center}
\caption{Subset of the posterior probability distribution for the frequency of the floating power spectrum coefficient in terms of $n$ such that the frequency is given by $\nu=n/T$.  Whilst several peaks are visible, only the inclusion of the dominant peak at $n=77.2$ results in an increase of the Evidence.}
\label{Figure:Freqposterior}
\end{minipage}
\end{figure}

\begin{figure}
\begin{minipage}{75mm}
\begin{center}$
\begin{array}{c}
\includegraphics[width=75mm]{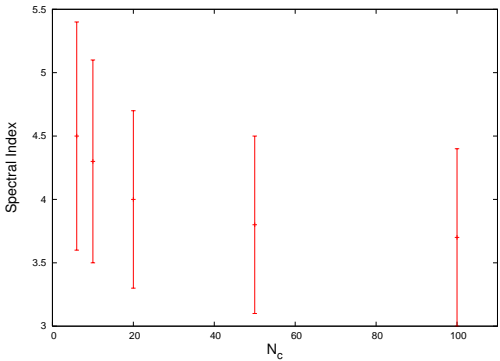} \\
\end{array}$
\end{center}
\caption{Mean spectral index and one sigma uncertainties for the power law model fitted to B1937+21 dataset 1.  Whilst the uncertainties are large there is a clear trend for the mean parameter estimate to move towards shallower spectral indices as the number of coefficients over which the model fit increases, and it intercepts the flat tail of the red spin noise spectrum.}
\label{Figure:SpectralIndices}
\end{minipage}
\end{figure}
Fig. \ref{Figure:Freqposterior} shows an section of the posterior probability distribution for the frequency of the floating power spectrum coefficient.  Whilst there are multiple modes in the posterior, only the inclusion of the dominant peak at a frequency of $n = 77.2/T$ is supported by the evidence. In total the optimal set of frequencies was given by  $[1,2,3,4,6,17.2,77.2]$.  Using this optimal set increases the log Evidence by $\sim 3$ over the model with 6 consecutive frequencies.  Fig. \ref{figure:D1Res} (left) shows the power spectrum for the red spin noise evaluated at the optimal set of frequencies, whilst the right panel shows the maximum likelihood signal realisation given the mean parameter estimates for this optimal set.   The power spectrum shows the clear signature of a steep red noise process at low frequencies, with a substantial, flatter, high frequency tail. 

When fitting a power law model to the data we find that the evidence stabilises after $N_c \sim 20$.  Despite the flat tail to the power spectrum visible with the model independent analysis, we observe an increase in the evidence of $\sim 1$ compared to the optimal frequency model.  Fig. \ref{Figure:SpectralIndices}   shows the mean posterior value and one-sigma uncertainties for the spectral indicies of the power law model fitted as we increase the number of coefficients included in the model.  Whilst the uncertainties are large, the mean value shows a clear trend towards flatter values decreasing from 4.5 to 3.7 as the number of coefficients included increases from 6 to 100 and more of the flat spectrum tail is included in the model.  

\begin{figure}
\begin{minipage}{75mm}
\begin{center}$
\begin{array}{c}
\includegraphics[width=75mm]{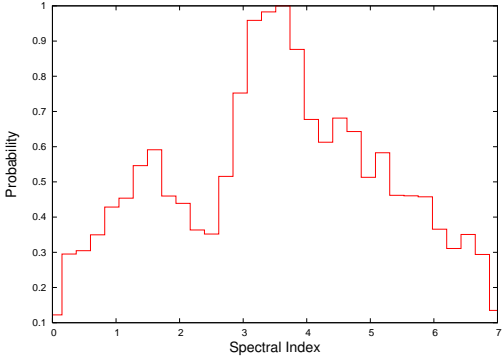} \\
\end{array}$
\end{center}
\caption{One dimensional marginalised posterior for the spectral indices of the two component power law model.  Two peaks are clearly visible in the posterior, corresponding to a steep red noise process at low frequencies, and a shallower one at higher frequencies.}
\label{Figure:Twospecs}
\end{minipage}
\end{figure}

Given these results we therefore also model the red noise as a two component power law. Fig. \ref{Figure:Twospecs} shows the one-dimensional marginalised posterior for one of the spectral indicies included in this model, displaying two clear peaks: one associated with the steep low frequency part of the spectrum with an index of $\sim 4$ and one with the flat spectrum tail with an index of $\sim 1.5$.  The evidence for the two component fit results in a final increase of $\sim$ 1 relative to the one component model, however we must subtract $\log2$ from this value in order to account for "counting degeneracy", the fact that we have two combinations of spectral indices (i.e. the parameters for the first and second power law can switch with out affecting the result).  Therefore, whilst there are tentative signs of this dual spectrum signal in the data, it is not sufficient to justify the additional parameters in our description of the stochastic signal, we therefore consider the one component power law to be the optimal choice for this dataset.

\subsubsection{Dataset 2}
\label{Section:1939D2}

\begin{table*}
\centering
\caption{Evidence for different stochastic models for pulsar B1937+21 in dataset 2} 
\centering 
\begin{tabular}{c c c} 
\hline\hline 
Model & $N_c$ (Red, DM) & $\log$ Evidence\\[0.5ex] 
\hline
\hline 
model independent red noise and DM with consecutive frequencies & 2 , 13 &  0\\
model independent red noise and DM with optimal frequencies & 5 , 12 &  4.3\\
power law red noise, model independent DM with consecutive frequencies & 10 , 13 & 0.7 \\
model independent red noise with consecutive frequencies, power law DM & 2 , 15 &   11.5\\
power law red noise and DM & 10 , 15 &  11.9\\
power law red noise and two component power law DM & 10 , 100 &  11.5\\
\hline
\end{tabular}
\label{Table:1937Ev2} 
\end{table*}

Table \ref{Table:1937Ev2} lists the evidence for the different models applied to dataset 2 where we give the value only for the optimal $N_c$ in each case.  In all cases the number of coefficients required by the DM variations was greater than for the red noise.  For dataset 2 the optimal set of frequencies is given by $[1,2,5,6]$ for the red spin noise, and $[1,2,3,4,6,8,10,11,12,13,24.6,81.5]$ for the DM variations.
We find that once again finding the optimal set of frequencies to include results in a significant increase in the $\log$ evidence of $\sim$ 4.5 when compared to the consecutive set.  As with dataset 1 we show in Fig \ref{figure:D1Res} the residuals and maximum likelihood signal realisation given the mean posterior values for the power spectrum coefficients obtained by the optimal model independent analysis.  Fitting for a power law model in both cases results in spectral indices of $-5 \pm 1$ and $-2.6 \pm 0.4$  for the red noise and DM variations respectively, consistent with the model independent analysis.  As with dataset 1 this consistency is supported by the log evidence, which has a maximum when constraining the power spectrum to follow a power law.

Compared to dataset 1 the we find neither the red noise nor the DM variation power spectrum in dataset 2 show any sign of a shallow tail.  This suggests that the DM model applied in K94 did not account fully for the higher frequency DM variation, flattening the red noise spectrum of the resultant residuals.

We now perform the analysis of the non-linear timing model simultaneously with the red noise and DM variations, where we model the latter two elements using a power law model as has been supported by the evidence in the preceding stochastic analysis.

\begin{figure*}
\begin{minipage}{180mm}
\begin{center}$
\begin{array}{cc}
\includegraphics[width=85mm]{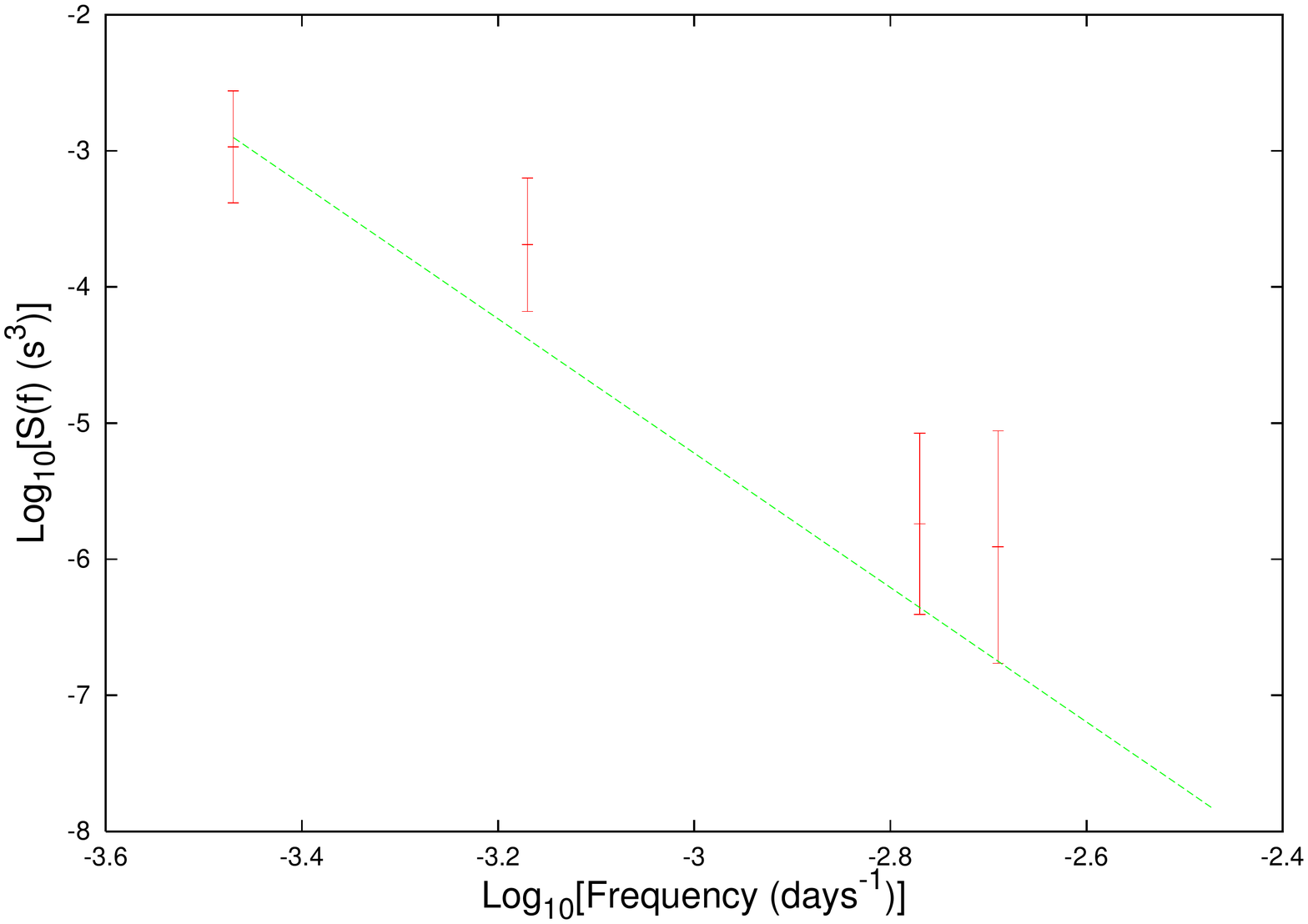} &
\includegraphics[width=85mm]{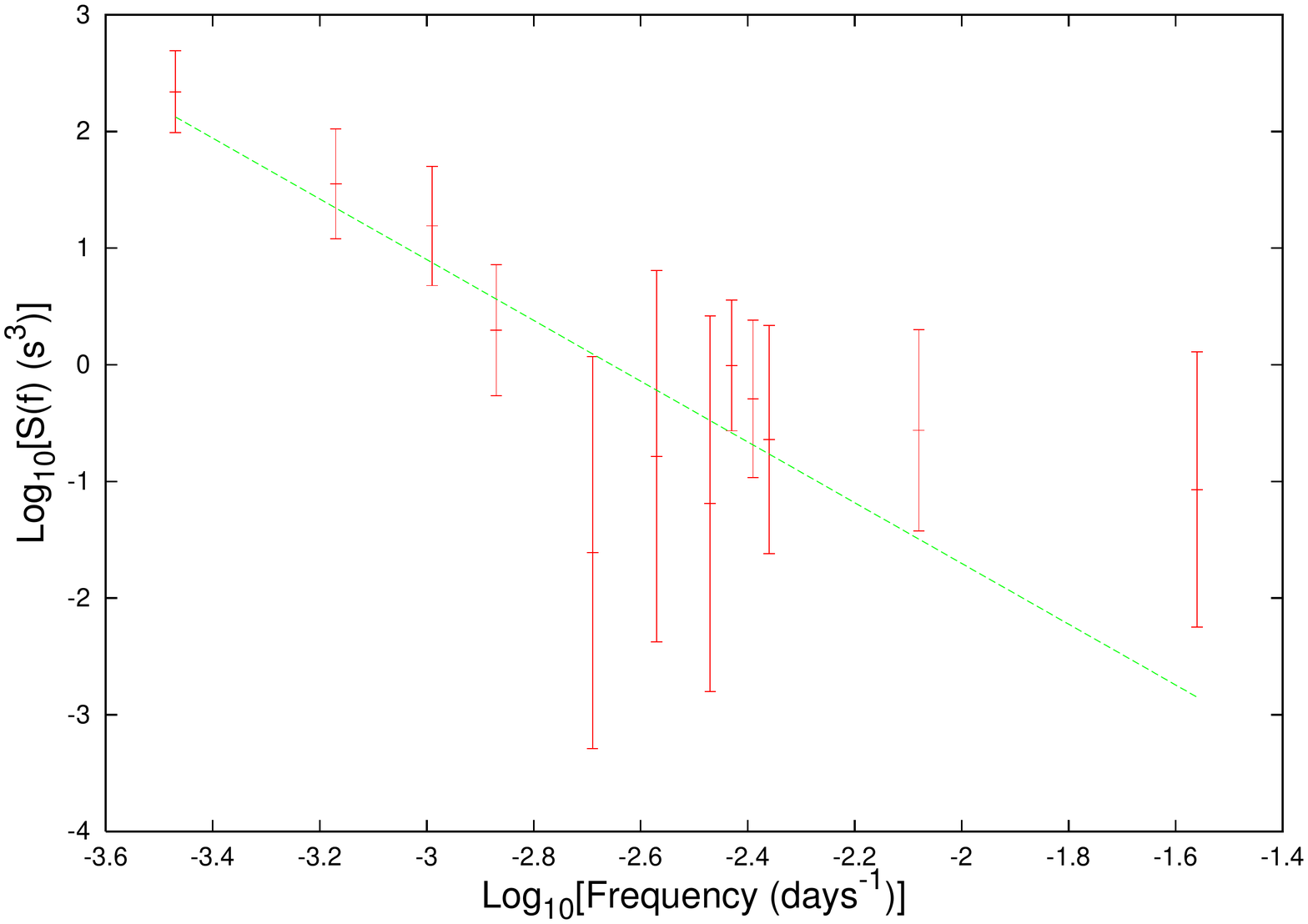} \\
\includegraphics[width=85mm]{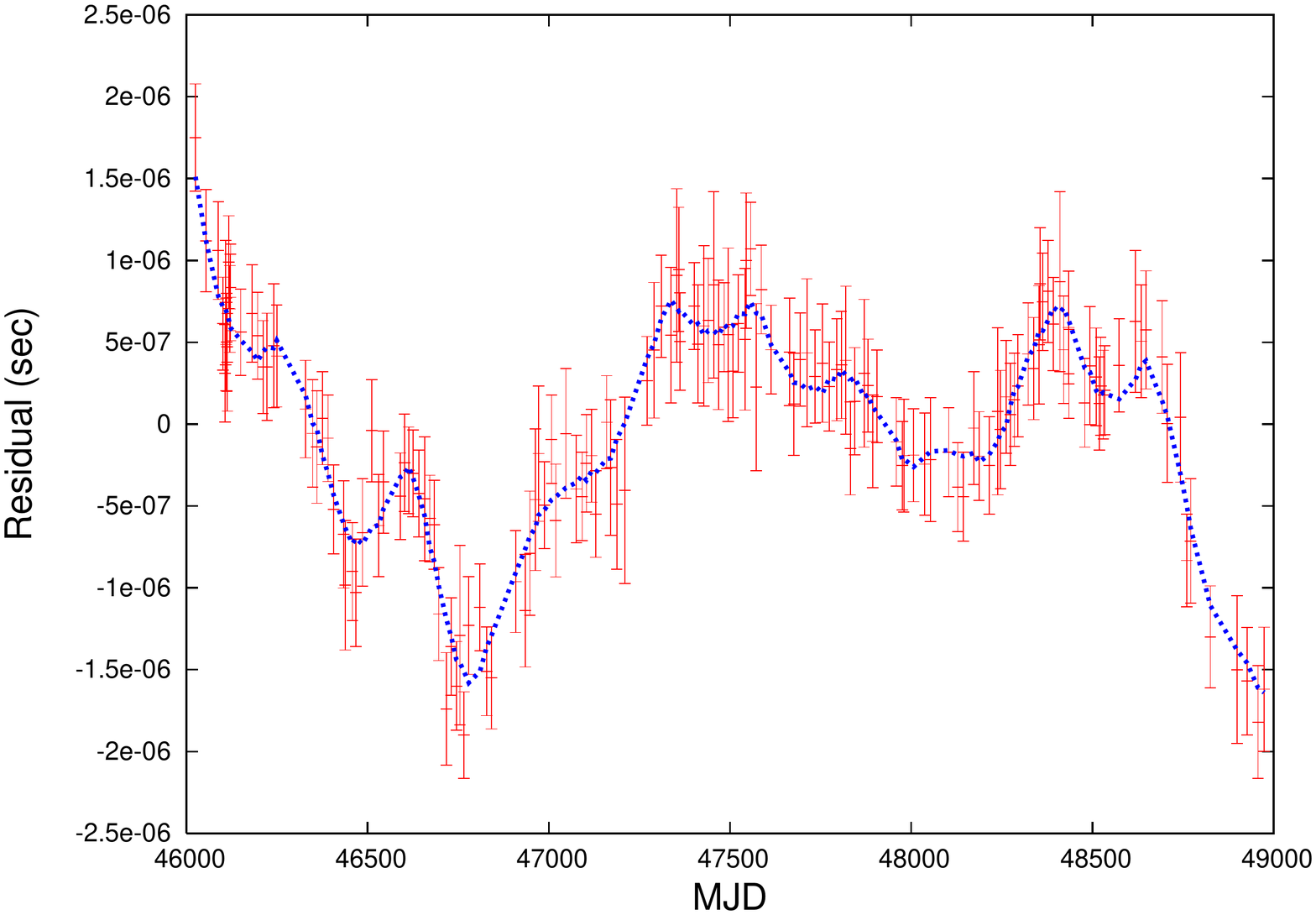} &
\includegraphics[width=85mm]{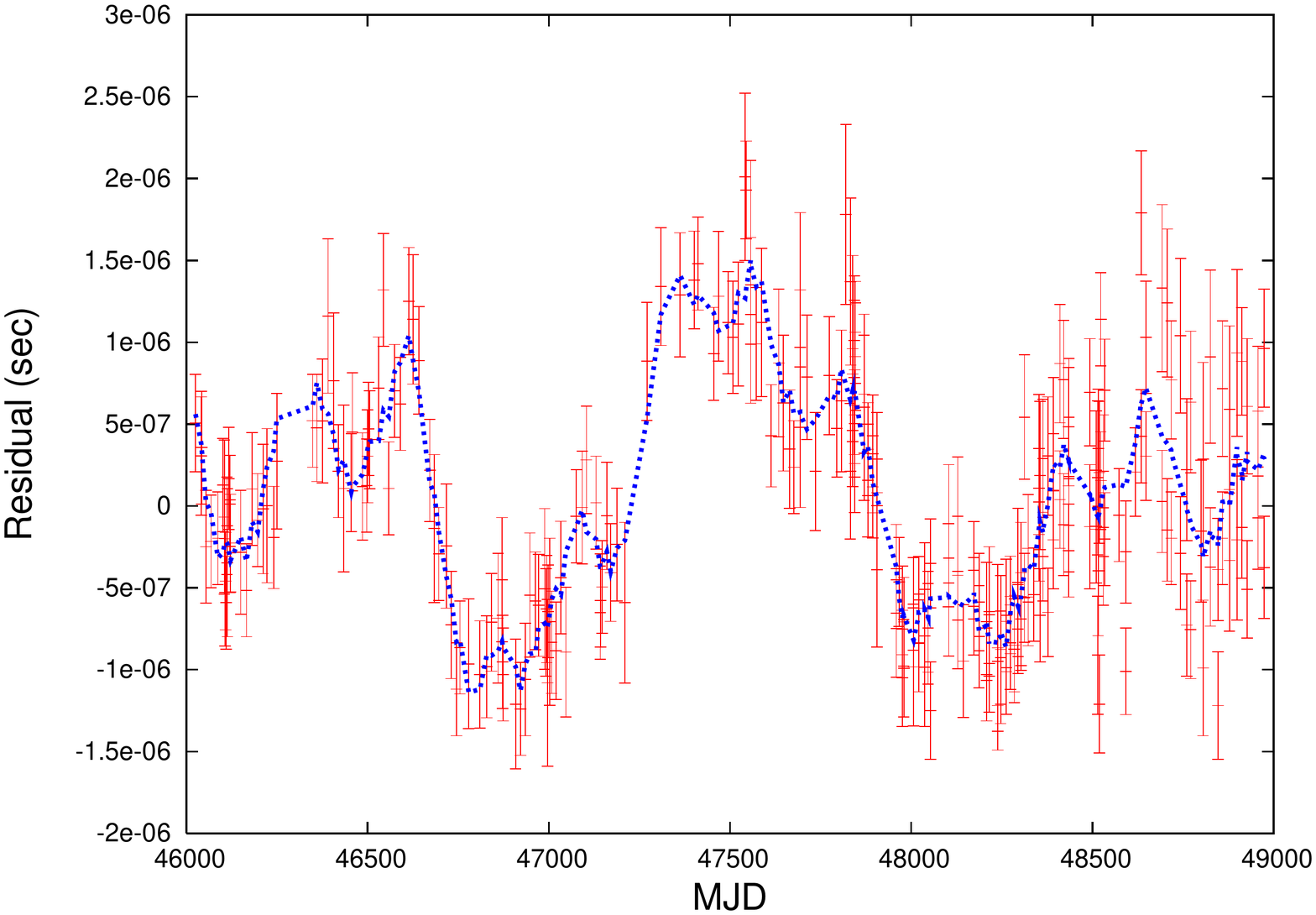} \\
\end{array}$
\end{center}
\caption{(top) Mean values and one sigma uncertainties for the optimal set of power spectrum coefficients required to describe the red noise (left) and DM variations (right).  The mean power law estimate is over plotted in both cases, for which we find spectral indices of $-5\pm1$ and $-2.6\pm0.4$ calculated including the first 10 and 15 consecutive coefficients for each respectively. (bottom) Timing residuals for B1937+21 dataset 2 (red points) and the best-fit combined signal realisation for the red noise and DM variations using the optimal model independent analysis (blue dotted line) shown for the 2.4GHz (left) and 1.4GHz (right) data separately for clarity. }
\label{figure:D1Res}
\end{minipage}
\end{figure*}

\subsubsection{Timing model analysis}

\begin{table*}
\caption{Timing model and stochastic parameter estimates for PSR B1937+21}
\begin{tabular}{lccc}
\hline\hline
\multicolumn{2}{c}{Fit and data-set} \\
\hline
Pulsar name\dotfill & B1937+21 \\ 
MJD range\dotfill & 46024.8---48973.8 \\ 
Number of TOAs\dotfill & 420 \\
\hline
\multicolumn{4}{c}{Measured Quantities - Including K94 DM Corrections} \\ 
\hline
& TempoNest & Tempo2 & Spectral Model \\
\hline
\hline
Right ascension, $\alpha$  \dotfill &  19:39:38.561314(10)& 19:39:38.561288(3) & 19:39:38.56130(5) \\ 
Declination, $\delta$ \dotfill & +21:34:59.1295(2)&  +21:34:59.13068(5) & +21:34:59.1291(9) \\ 
Pulse frequency, $\nu$ (s$^{-1}$)\dotfill & 641.92823355803(16) & 641.9282335579857(7) & 641.9282335581(14)\\ 
First derivative of pulse frequency, $\dot{\nu}$ (s$^{-2}$)\dotfill & $-$4.33169(3))$\times 10^{-14}$& $-$4.3316913(15)$\times 10^{-14}$  & $-$4.33169(5)$\times 10^{-14}$\\ 
Dispersion measure, $DM$ (cm$^{-3}$pc)\dotfill & 71.04003(2)& 71.039981(15) & 71.040024(15)\\ 
Proper motion in right ascension, $\mu_{\alpha}$ (mas\,yr$^{-1}$)\dotfill & 0.084(11)& 0.054(4) & 0.08 (7) \\ 
Proper motion in declination, $\mu_{\delta}$ (mas\,yr$^{-1}$)\dotfill &$-$0.421(15) & $-$0.319(4) & $-$ 0.43(2) \\ 
Parallax, $\pi$ (mas)\dotfill & 0.24(9)& 0.01(4)  & 0.20(15)\\ 
EFAC\dotfill &0.88(14) & -- & --\\
$\log_{10}$ EQUAD\dotfill & $-$6.59(4)& -- & --\\
$\log_{10} \mathrm{A_{red}}$ \dotfill &  $-$3.30(16) & -- & --\\
$\gamma_{\mathrm{red}}$ \dotfill & 3.9(6) & -- & -- \\
\hline
\multicolumn{4}{c}{Measured Quantities - Excluding K94 DM Corrections} \\ 
\hline
& TempoNest & Tempo2 & Spectral Model\\
\hline
\hline
Right ascension, $\alpha$  \dotfill & 19:39:38.561307(10) & 19:39:38.561219(3) &  19:39:38.561301(19) \\ 
Declination, $\delta$ \dotfill  & +21:34:59.1296(2) & +21:34:59.13136(5) &  +21:34:59.1292(4)\\ 
Pulse frequency, $\nu$ (s$^{-1}$)\dotfill & 641.9282335581(3) & 641.9282335580713(13) & 641.9282335581(9) \\ 
First derivative of pulse frequency, $\dot{\nu}$ (s$^{-2}$)\dotfill &$-$4.33169(6) $\times 10^{-14}$& $-$4.331679(3)$\times 10^{-14}$ & $-$4.3317(2)$\times 10^{-14}$\\ 
Dispersion measure, $DM$ (cm$^{-3}$pc)\dotfill & 71.041(13) & 71.040715(15) & 71.04060(12)\\ 
Proper motion in right ascension, $\mu_{\alpha}$ (mas\,yr$^{-1}$)\dotfill & 0.069(10) & $-$0.005(3) &0.08(2) \\ 
Proper motion in declination, $\mu_{\delta}$ (mas\,yr$^{-1}$)\dotfill &$-$0.411(15) & $-$0.279(4) & $-$0.44(5)\\ 
Parallax, $\pi$ (mas)\dotfill & 0.25(10) & 0.41(4) & 0.19(13)\\ 
EFAC\dotfill &1.29(19) & -- & --\\
$\log_{10}$ EQUAD\dotfill & $-$7.3(9)& -- & --\\
$\log_{10} \mathrm{A_{red}}$ \dotfill &  $-$3.7(3) & -- & --\\
$\gamma_{\mathrm{red}}$ \dotfill & 5.3(9) & -- & --\\
$\log_{10} \mathrm{A_{DM}}$ \dotfill & $-$0.15(7) & -- & --\\
$\gamma_{\mathrm{DM}}$ \dotfill & 2.7(3) & --  & --\\
\hline
\multicolumn{2}{c}{Set Quantities} \\ 
\hline
Epoch of frequency determination (MJD)\dotfill & 52601 \\ 
Epoch of position determination (MJD)\dotfill & 52601 \\ 
Epoch of dispersion measure determination (MJD)\dotfill & 52601 \\ 
\hline
\multicolumn{2}{c}{Assumptions} \\
\hline
Clock correction procedure\dotfill & TT(TAI) \\
Solar system ephemeris model\dotfill & DE405 \\
Binary model\dotfill & NONE \\
Model version number\dotfill & 5.00 \\ 
\hline
\end{tabular}
\label{Table:B1937results}
\end{table*}

\begin{figure*}
\begin{minipage}{168mm}
\begin{center}$
\begin{array}{c}
\hspace{-2.0cm}
\includegraphics[width=200mm]{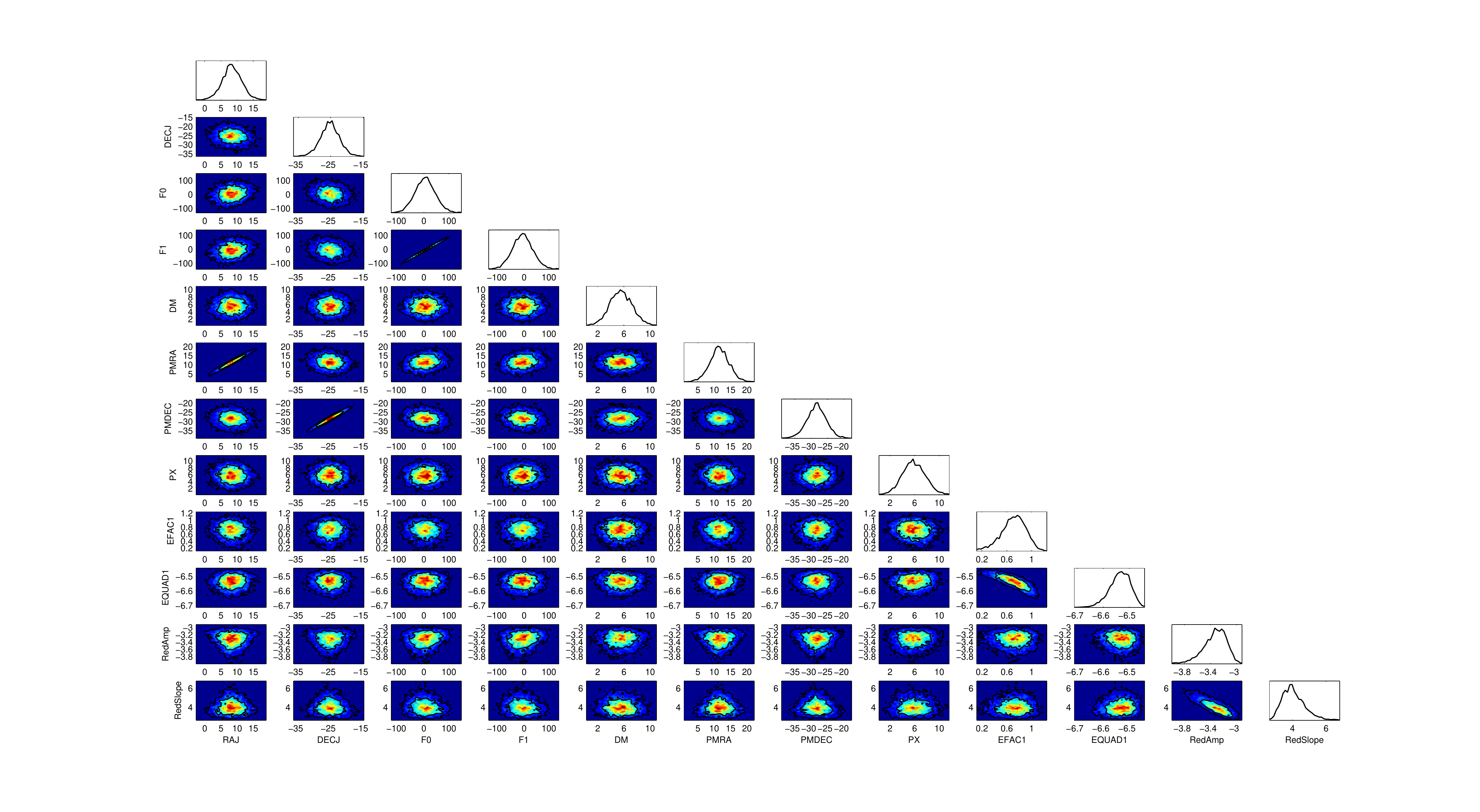} \\
\hspace{-2.0cm}
\includegraphics[width=200mm]{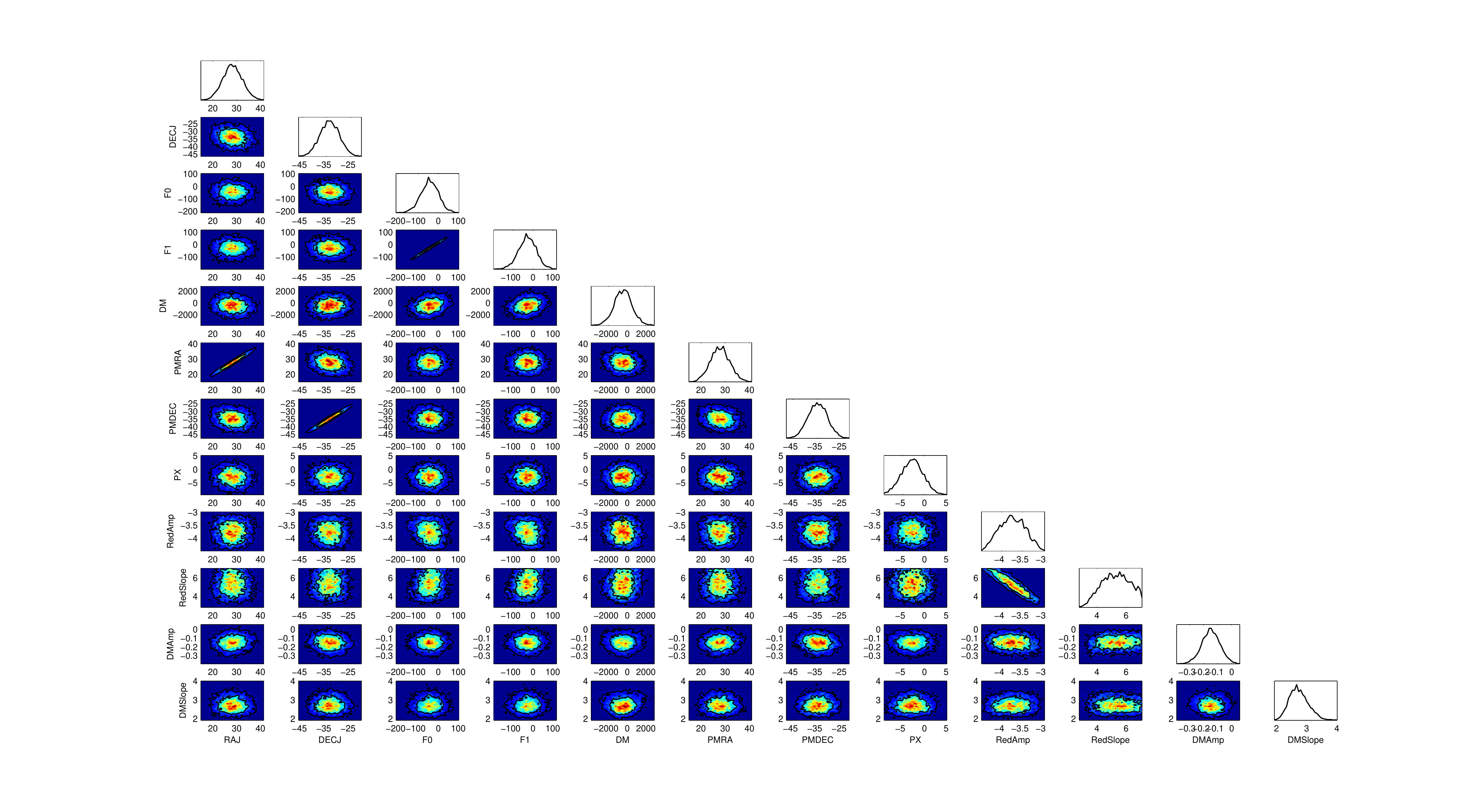} 
\end{array}$
\end{center}
\caption{(top) The one and two dimensional marginalised posterior distributions for the eight timing model parameters: RA, DEC, F0, F1, DM, PMRA, PMDEC and PX, and the four stochastic parameters: EFAC, EQUAD, and amplitude and spectral index of the power law model for the red noise for dataset 1.  (bottom) As in the top plot, but for dataset 2, and with EFAC and EQUAD replaced with the amplitude and spectral index of the power law model for dispersion measure variations.}
\label{figure:DMPlots}
\end{minipage}
\end{figure*}

Table \ref{Table:B1937results} lists the mean posterior values and associated one sigma uncertainties for our final timing model and stochastic solutions to both datasets 1 and 2 where, following the results in Sections \ref{Section:1939D1} and \ref{Section:1939D2}, the stochastic signals have been modelled as power law processes for both the red noise and DM variations.  Fig. \ref{figure:DMPlots} then shows the one and two dimensional marginalised posteriors for a selection of the modelled parameters; RA, DEC, PMRA, PMDEC, PX and the red noise and DM spectral indices and amplitudes.  As expected the timing model parameters show no evidence for non-linear behaviour in either case despite the high levels of red noise in the dataset.  As such we would expect that our estimates for the stochastic parameters when analytically marginalising over the timing model will be completely consistent with those in the full analysis and indeed this is the case in both datasets. 

Table \ref{Table:B1937results} also lists the timing model solutions returned by applying both the standard Tempo2 fit and using the SpectralModel plug-in which utilises the Cholesky method described in C11.   In order to use the SpectralModel plug-in in the case of dataset 2 we therefore applied the method of \cite{2013MNRAS.429.2161K} to model the DM variations as a linearly interpolated time series, sampled every 100 days in the observation.  Unlike the approach in K94 this therefore allows us to include the uncertainties in the DM model in the final timing model solutions whilst still using the Cholesky method.

\begin{figure*}
\begin{minipage}{168mm}
\begin{center}$
\begin{array}{cc}
\hspace{-2.0cm}
\includegraphics[width=100mm]{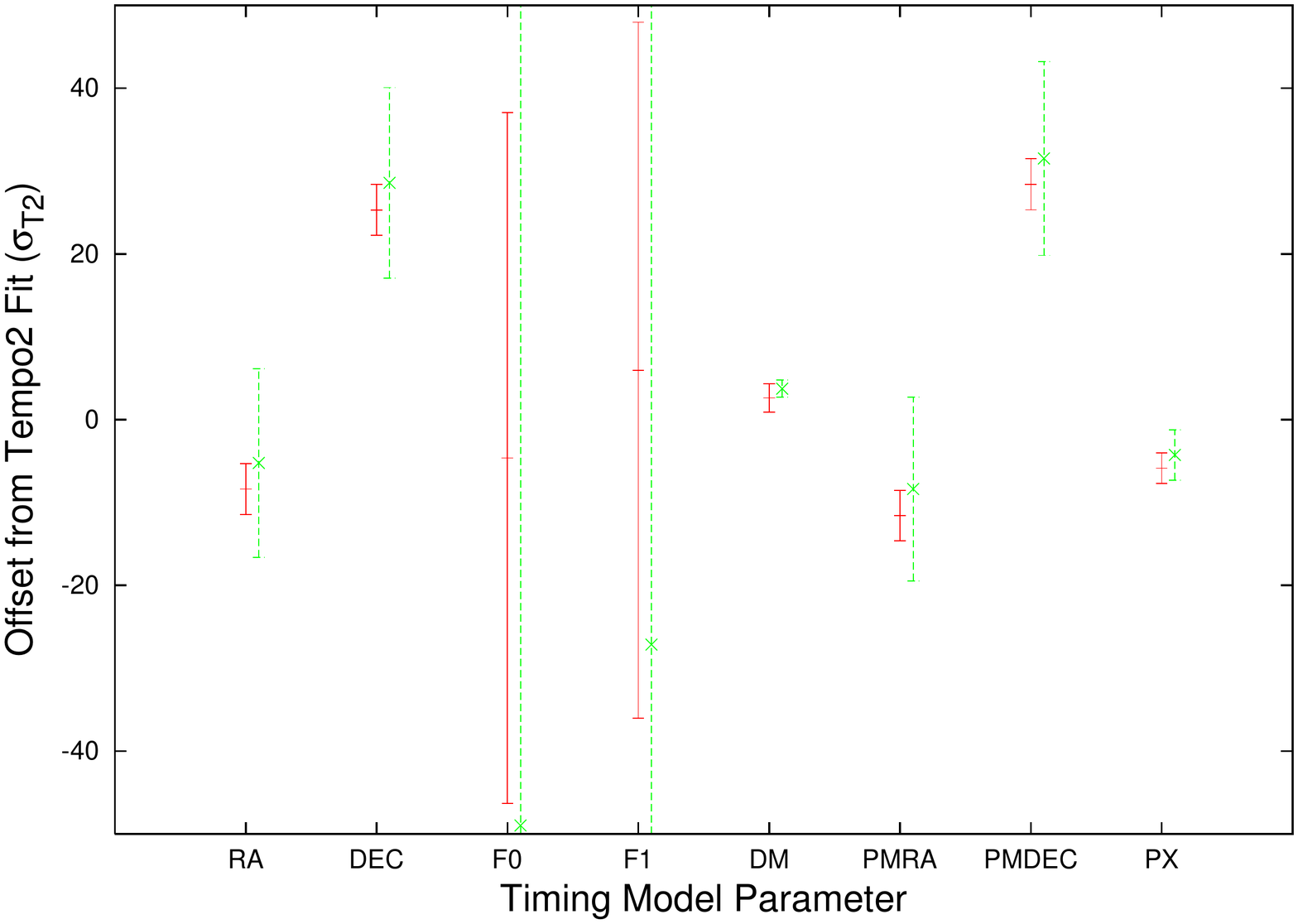} &
\hspace{-1.0cm}
\includegraphics[width=100mm]{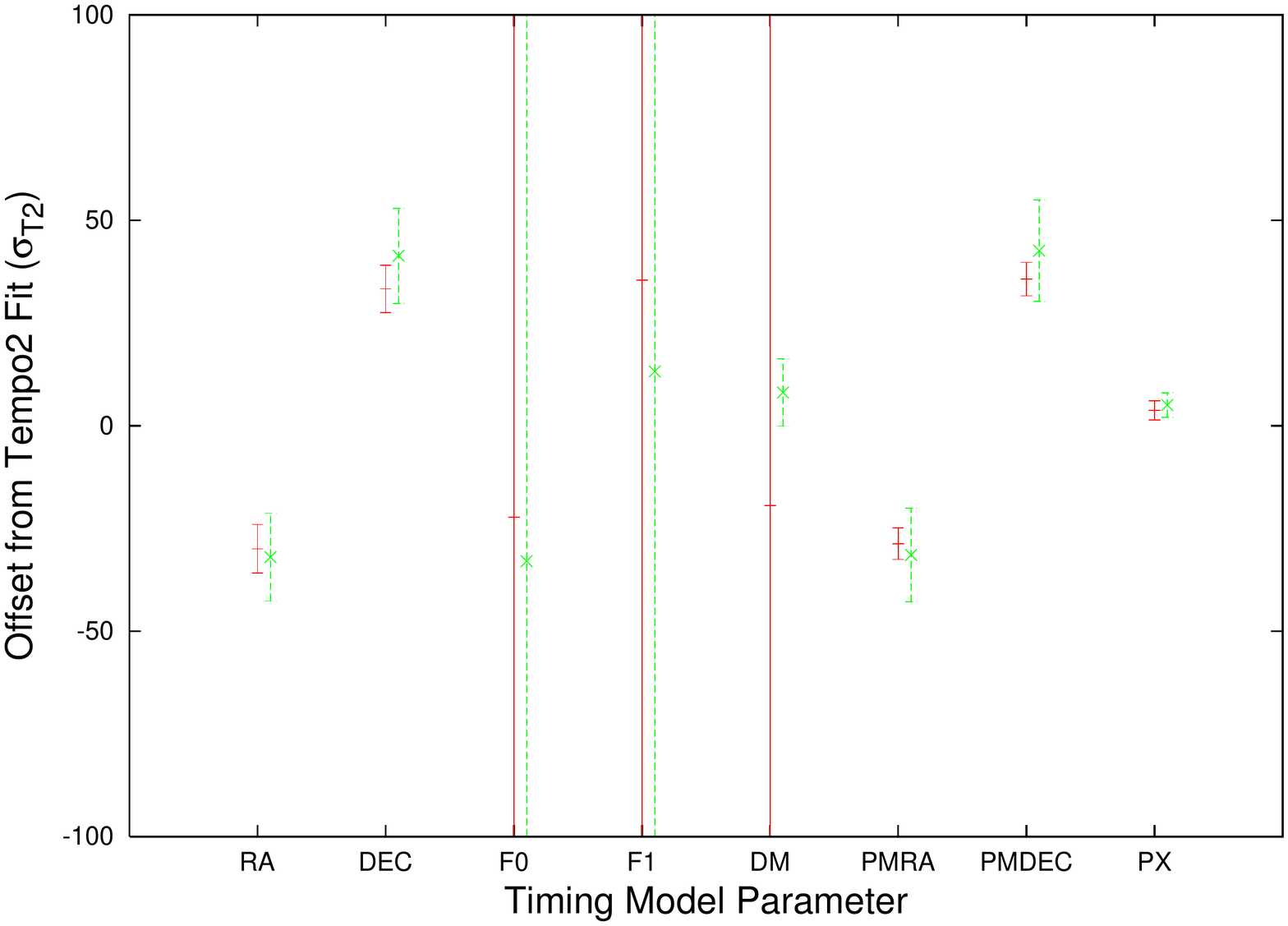} \\
\end{array}$
\end{center}
\caption{(left) Difference between the parameter estimates and uncertainties returned by TempoNest (red solid points) and the SpectralModel plug-in (blue points) and Tempo2 for PSR B1937+21 for dataset 1.   (right) As with the left plot, but for dataset 2, where we have used DMModel to characterise the dispersion measure variations for the SpectralModel fit, and a power law DM model for the TempoNest fit.  In both plots the zero on the $y$ axis represents the value returned by Tempo2, with the $y$ axis being in units of the 1$\sigma$ Tempo2 error.  Therefore, a fit consistent with that returned by Tempo2 will have a value of $0\pm1$ in these plots.}
\label{figure:DMTime}
\end{minipage}
\end{figure*}

Fig. \ref{figure:DMTime} shows a graphical representation of the differences between the TempoNest, SpectralModel and Tempo2 results for the timing model parameters given in Table \ref{Table:B1937results} for datasets 1 (left) and 2 (right).  Here the value on the yaxis is given by the difference between the TempoNest (or SpectralModel) estimate, and the Tempo2 estimated parameter values, divided by the Tempo2 error, which as before we will denote $\sigma_{T2}$.  The errors are then the 1$\sigma$ errors estimated by TempoNest (SpectralModel) which we will denote $\sigma_{TN}$ ($\sigma_{SM}$).  Therefore a fit that is consistent both with the value and error returned by Tempo2 would have a value of $0\pm1$ in Fig. \ref{figure:DMTime}.

There are several things that are immediately apparent in Fig. \ref{figure:DMTime}.  The first is the level of disparity between the parameter estimates returned by Tempo2, and those returned by both TempoNest and the SpectralModel analysis.  With the exception of only DM in dataset 1, and PX in dataset 2 the parameter estimates returned by TempoNest during the joint analysis are at least 4$\sigma_{T2}$ away from the Tempo2 values, with an average deviation of $\sim 30 \sigma_{T2}$.

Whilst the parameter estimates returned by the TempoNest and SpectralModel analysis are in both cases consistent with one another, the estimates of the uncertainties are  several times larger than those returned by TempoNest by a factor of $\sim 2-3$.

In order to investigate the difference in the uncertainties returned by TempoNest and the SpectralModel plugin, we first consider the major differences between the different types of analysis:

\begin{itemize}
\item TempoNest uses the full non-linear model compared to the linear approximation in SpectralModel
\item TempoNest uses physical priors on parameters such as parallax, such that they must take values greater than zero, SpectralModel does not
\item TempoNest includes both red and white noise estimation in the fit with the timing model, SpectralModel does not include additional white noise terms, and fixes the red noise when calculating the uncertainties for the timing model.
\item TempoNest uses a cut off power law model for the red noise, whereas SpectralModel uses a red noise model with a corner frequency at which it turns over.
\end{itemize}
In order to ascertain how large an effect these differences make we therefore rerun the analysis on dataset 1, so that we have the same DM model in both cases, whilst making the following changes:
\begin{description}
\item[1:] Use the linear timing model, linearised at the timing model parameters estimated by the SpectralModel plug-in without the condition that parameters must take physical values
\item[2:] As in 1, but fixing $EFAC=1$ and $EQUAD=0$, and fixing the spectral index of the red noise power law to be the same value used in the SpectralModel plug-in ($-4.0$)
\end{description}
\begin{figure}
\begin{center}$
\begin{array}{c}
\includegraphics[width=80mm]{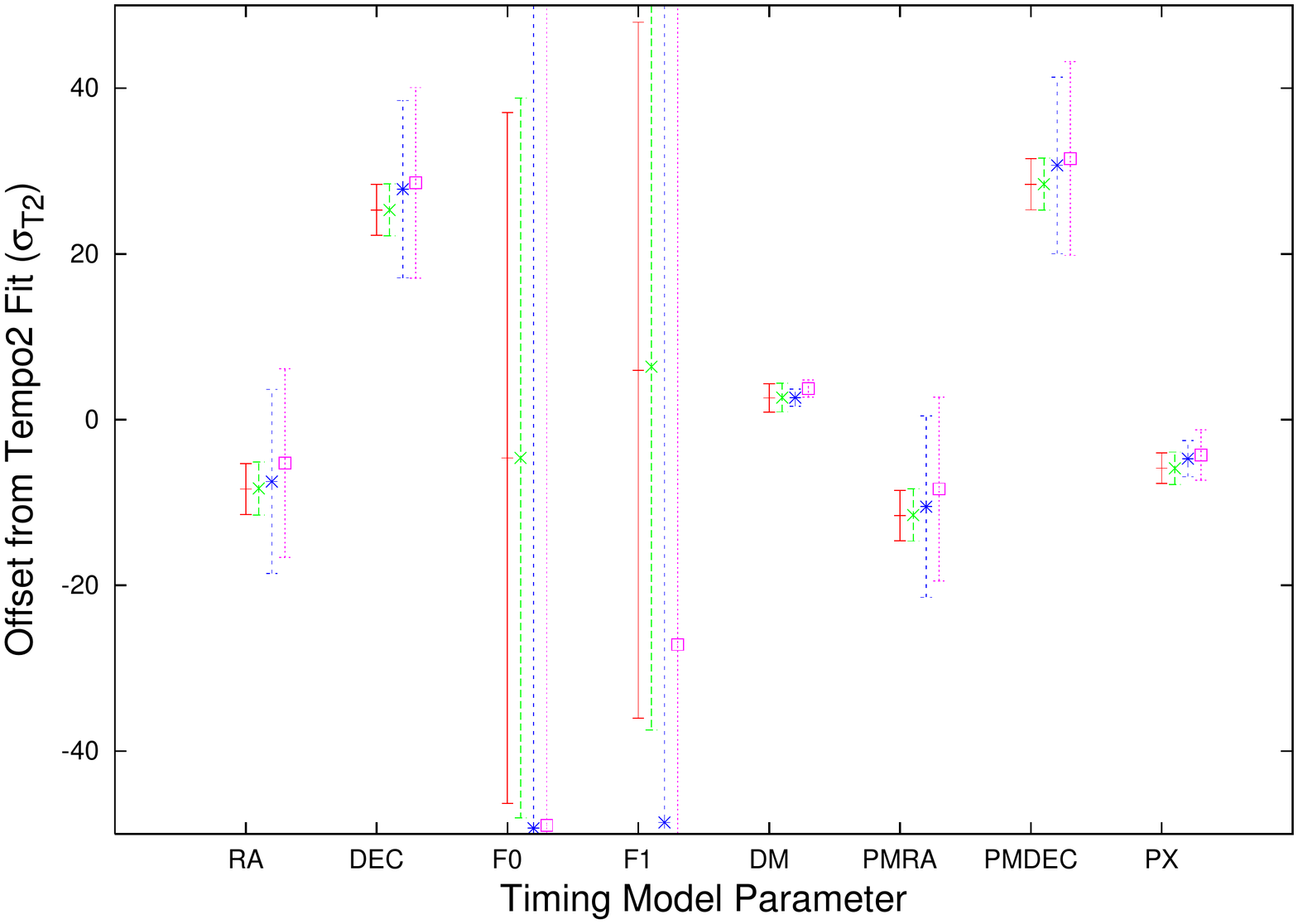} \\
\end{array}$
\end{center}
\caption{Difference between the timing model parameter estimates returned by TempoNest under a series of different approximations and the Tempo2 parameter estimates in the same format as Fig. \ref{figure:DMTime}, as well as the timing model estimates returned by the SpectralModel plug-in for dataset 1.  For each timing model parameter from left to right the data points represent the following: i )Full analysis using TempoNest ii) Using the linear timing model, linearised at the timing model parameters estimated by the SpectralModel plug-in, iii) as (ii) but with the $EFAC=1, EQUAD=0, \gamma_{\mathrm{red}}=-4.0$, and, iv) the SpectralModel fit}
\label{figure:FinalComp}
\end{figure}
Fig. \ref{figure:FinalComp} shows the timing model parameters estimated from these different approximations in the same format as in Fig. \ref{figure:DMTime}.  The red and green points represent the parameter estimates from the full non-linear timing analysis, and the linearised analysis respectively, however there is no discernible difference in either the estimated parameter values, or the uncertainties between these two models, as expected given the set of parameters included in the timing model.  The blue points represent the case where we have not included white noise in the model, and have fixed the spectral index of the red noise to be that used in the SpectralModel analysis ($-4.0$).  Here there is a clear increase in the level of uncertainty of the timing model parameters relative to the full analysis by a factor of $\sim 2-3$ except for DM which sees the uncertainties decrease by $\sim 60\%$.  In both cases however this brings the estimated uncertainties in line with those derived from the SpectralModel analysis, with remaining differences being of the order $10\%$, with the exception of the quadratic spin down parameters, from which we would expect the greatest difference given the use of different red noise models.

This clearly emphasises the importance of including as much as is required to fully describe the data simultaneously in the fit along with the timing model parameters.  We should make clear that the magnitude of the differences observed in PSR B1937+21 will not be typical for most millisecond pulsars, however, the precise level of difference to expect is difficult to quantify a priori, being a function of signal-to-noise, the cadence of the observations, and the complexity of the timing model used to describe the pulsar.  We therefore still suggest that unless a full simultaneous analysis such as that described here is performed in every case, the unpredictable variation in the uncertainties returned from the analysis must impact the robustness of the science extracted from that analysis.

\section{Conclusion}
\label{Section:Conclusions}

We have introduced TempoNest, a software package that provides a means of performing a robust Bayesian analysis of pulsar timing data.  TempoNest allows for the joint analysis of the timing model along with a range of additional stochastic parameters including EFAC and EQUAD parameters, and descriptions of both red spin noise and dispersion measure variations using either a power law description of the noise, or in a model-independent way, parameterising the power at individual frequencies in the data.

We have applied TempoNest to both simulated and real datasets in order to demonstrate several key aspects of functionality.  First we used simulated data for the pulsar PSR J1713+0747 in order to compare the linear and non-linear timing models where the level of noise varied between simulations, from that expected from the square kilometer array ($\sim$ 100ns white timing noise), to a level more representative of current observations, including red spin noise.  We showed that in the high signal--to--noise example the differences between the timing solutions for the linear and non-linear timing model parameters were negligible.  In the lower signal--to--noise examples, however, the linear timing model failed to capture all the information present in the data, with large curving degeneracies in the non-linear parameter space leading to uncertainties that exceeded those in the linear approximation by up to an order of magnitude.

We then applied TempoNest to two publicly available datasets, the binary pulsar B1855+09 and the isolated pulsar B1937+21.  For the former we used a model independent method of parameterising the red spin noise in the data and found marginal support for a single low frequency component in addition to the timing model parameters fitted, but found it did not affect the timing model solutions in any observable way, with parameter estimates that were completely consistent with those of Tempo2 in all respects.  We then demonstrated the ability for TempoNest to perform model selection using the evidence by including a series of additional timing model parameters and repeating the analysis in order to find the optimal set that described the data.  This included adding jumps between instrument back ends, and additional physical parameters such as derivatives of the binary period or eccentricity.  

In the case of B1937+21 we used both a power law, and model independent method of parameterising both the dispersion measure and red noise signals in the data, and found whilst they gave consistent results, the evidence heavily favoured the use of the simpler model, with both components being well described by a power law power spectrum with spectral indices of $-2.7\pm 0.3$ and $-5.3 \pm 0.9$ respectively.  When comparing the timing model solutions returned by TempoNest from this joint analysis with those of Tempo2 we found large discrepancies, both in terms of the parameter estimates themselves and their uncertainties.  In the most extreme cases the TempoNest parameter estimates were up to $\sim 38 \sigma_{T2}$ away from the Tempo2 values, with  $\sigma_{T2}$ the returned Tempo2 uncertainty, whilst the uncertainties themselves were over two orders of magnitude greater in the case of the pulsar's spin down parameters.  When compared to the Cholesky method found in the Tempo2 SpectralModel plug-in, we found that by not including all the stochastic processes in the analysis simultaneously with the timing model, the timing model parameter uncertainties are overestimated by a factor $\sim 2-3$ in almost all cases, showing unambiguously the importance of including as much as is required to fully describe the data simultaneously in the analysis.  This is all the more critical given the precise level of difference to expect for any pulsar is difficult to quantify a priori, being a function of signal-to-noise, the cadence of the observations, and the complexity of the timing model used to describe the pulsar.  We therefore suggest that unless a full simultaneous analysis such as that described here is performed in every case, the unpredictable variation in the uncertainties returned from the analysis must impact the robustness of the science extracted from that analysis.

TempoNest is freely available as a development build\footnote{https://github.com/LindleyLentati/TempoNest}, with a full public release planned in the near future.

\label{lastpage}

\end{document}